%% file: els-article.tex
\documentclass[onecolumn,authoryear]{els-mrw} 

\usepackage{amsmath,amssymb,amsfonts,amsthm,makeidx,graphicx}
\usepackage{txfonts}
\usepackage{helvet}

\usepackage{hyperref}
\usepackage{longtable}

\usepackage{booktabs}
\usepackage{multirow}
\usepackage{pdflscape}
\usepackage{threeparttablex}

\hypersetup{
    colorlinks=true,
    linkcolor=blue,
    filecolor=magenta,      
    urlcolor=cyan,
    citecolor=blue,
    pdfpagemode=UseThumbs,
    }
    
\newcommand\aap{A\&A}                
\newcommand\aapr{A\&ARv}             
\newcommand\aj{AJ}                   
\newcommand\apj{ApJ}                 
\newcommand\apjl{ApJ}                
\newcommand\apjs{ApJS}               
\newcommand\apss{Ap\&SS}             
\newcommand\araa{ARA\&A}             
\newcommand\gca{Geochimica Cosmochimica Acta}   
\newcommand\mnras{MNRAS}             
\newcommand\na{New~Astron.}          
\newcommand\pra{Phys. Rev.~A}        
\newcommand\prd{Phys. Rev.~D}        
\newcommand\prl{Phys. Rev.~Lett.}    
\newcommand\pasp{PASP}               
\newcommand\physrep{Phys.~Rep.}      
\newcommand\solphys{Sol.~Phys.}      
\newcommand\ssr{Space Sci. Rev.}     
\newcommand\zap{Z.~Astrophys.}       

\begin{document}

\chapter{The Chemical Composition of the Sun}\label{chap1}

\author[1]{Maria Bergemann}%
\author[2]{Katharina Lodders}%
\author[3]{Herbert Palme}%

\address[1]{\orgname{Max Planck Institute for Astronomy}, \orgaddress{Koenigstuhl 17, 69117, Heidelberg, Germany}}
\address[2]{\orgname{Washington University}, \orgdiv{Department of Earth, Environmental, \& Planetary Sciences and McDonnell Center for Space Sciences}, \orgaddress{St. Louis, MO 63130, USA}}
\address[3]{\orgname{Senckenberg Forschungsinstitut und Naturmuseum}, \orgaddress{Frankfurt, Germany}}

\articletag{Chapter Article tagline: new article}

\maketitle

\begin{glossary}[Glossary]

\end{glossary}

\begin{glossary}[Nomenclature]
\begin{tabular}{@{}lp{34pc}@{}}
atwt(i) & atomic weight of element $i$\\
A    & atomic mass number (not italic)\\
Z    & atomic number (not italic)\\
A(i) & abundance of the chemical element $i$ in the astronomical abundance scale\\
$X,Y,Z$ & mass fraction of H, He, and metals, respectively (italic) \\
1D  & 1-dimensional\\
3D  & 3-dimensional\\
LTE & Local Thermodynamic Equilibrium\\
NLTE & not in Local Thermodynamic Equilibrium (statistical equilibrium or time-dependent rate equations) \\
RHD & Radiation-HydroDynamics\\
SSM & Standard Solar Model\\
UV & ultra-violet\\
IR & infra-red\\
SF & Scaling Factor\\
IUPAC & International Union of Pure and Applied Chemistry\\
CME & Coronal Mass Ejections\\
CI chondrites & C for carbonaceous and I for \textit{Ivuna} a type area in Tanzania; CI-type carbonaceous chondrites \\
ppm & parts-per-million\\
atwt(i) & atomic weight of the chemical element $i$\\

\end{tabular}
\end{glossary}

\begin{abstract}[Abstract]
This chapter provides a brief introduction to the chemical composition of the Sun. The focus of the chapter is on results obtained from the physical analysis of the solar photosphere. Data obtained from meteorites, solar wind and corona measurements, as well as helioseismology, and solar neutrinos are briefly reviewed. The elemental and isotopic composition of the solar system is derived by combining the solar and meteoritic data. The cosmochemical and astronomical abundance scales are described. The results of the determinations of the protosolar chemical composition, as well as the initial and present-day mass fractions of hydrogen, helium, and metals ($X,Y,Z$) for the solar system are presented in extensive tables. All tables are also  available in machine-readable form via Zenodo \url{https://doi.org/10.5281/zenodo.14988840}
\end{abstract}

\section{Key points}\label{s1}%

\begin{itemize}
\item Solar photosphere abundances can be measured using solar photosphere models for 60 chemical elements from Li (Z$=$3) to Th (Z$=$90); direct fully model-independent measurements are not possible;
\item These data represent the most robust and diverse set of chemical composition measurements in astronomy;
\item All elements are measured relative to hydrogen (H); 
\item H and noble gases (He, Ne, Ar, Kr) are not measurable in the solar photospheric spectra; Xe and Kr are based on interpolation of s-process nuclide abundances using the s-process model and Galactic chemical evolution yields; 
\item Spectroscopy of sunspots provide abundance estimates for F, Cl, In, and Tl; these model-dependent estimates currently rely on 1D LTE models. Assumptions have to be made on the structure of sunspots and on magnetic field strength and orientation; 
\item Different instruments, including those on ground- and space-based astronomical facilities, can be used to obtain high-resolution solar spectra; 
\item Physical models used for solar photospheric abundance determinations include realistic gas dynamics, departures from local thermodynamic equilibrium (NLTE), and radiative transfer in 3-dimensional geometry; 
\item So far, 3D NLTE models have been used only for a dozen of chemical elements (incl. C, N, O, Si, Mg, Mn, Fe, Ba, Y, Eu); for these elements, the error of abundances is typically around or better than 0.1 dex (26$\%$);
\item The photospheric abundances can be contrasted with indirect measurements based on solar neutrino fluxes (sum of C and N), solar wind and corona (including Mg, Fe, Ca), CI chondritic meteorites (majority of elements, except the most volatile elements and noble gases), and analysis of the solar interior structure (C,N,O);
\item The comparison between atmospheric and interior composition is limited by the knowledge of gravitational settling corrections (model-based and assumed to be around 10-20$\%$); 
\item For most chemical elements, the agreement between CI meteorites and the photospheric abundances is excellent; there is no strong evidence for a volatility trend between the solar photospheric and CI chondritic data;
\item The isotopic abundances are mainly based on solar wind measurements and they are available only for selected elements (incl. C, N, O);
\item Recent 3D NLTE measurements by two independent groups arrive at different conclusions, regarding the key chemical elements and bulk solar metallicity. The low-metallicity solar abundances \citep{Asplund2021} present a problem for the models of the solar interior. In contrast, the high-metallicity solar abundances \citep{Bergemann2021, Magg2022} lead to largely consistent predictions of the Standard Solar Models and helioseismology. These values are consistent with measurements based on solar neutrino fluxes, and with combined analyses of solar wind and solar system data. 

\end{itemize}

\section{Introduction}\label{chap1:sec1}

The chemical composition of the Sun is a key parameter in modern astrophysics. Solar element abundances are used as input parameters in models attempting to understand the physical properties, internal structure, and evolution of the Sun and other stars. The knowledge of the Sun’s composition is also important for modeling the formation and evolution of the solar system and its planets. Solar abundances are also taken as a baseline for comparison with chemical abundances of other astronomical objects, including extrasolar planets, circumstellar and protoplanetary discs, interstellar medium, stellar clusters, and galaxies. Ideas on the origin of the elements and models of the stellar nucleosynthesis \citep[e.g.][]{Goldschmidt1938, Suess1947a, Suess1947b, Burbidge1957, Cameron1957} were originally motivated by the relative abundances of elements in the Earth’s crust but quickly shifted to the solar photosphere. Such models use the solar abundances as a starting point to understand element and energy production in stars during their various stages of evolution. 

\begin{figure}[t]
\centering
\includegraphics[width=0.8\textwidth]{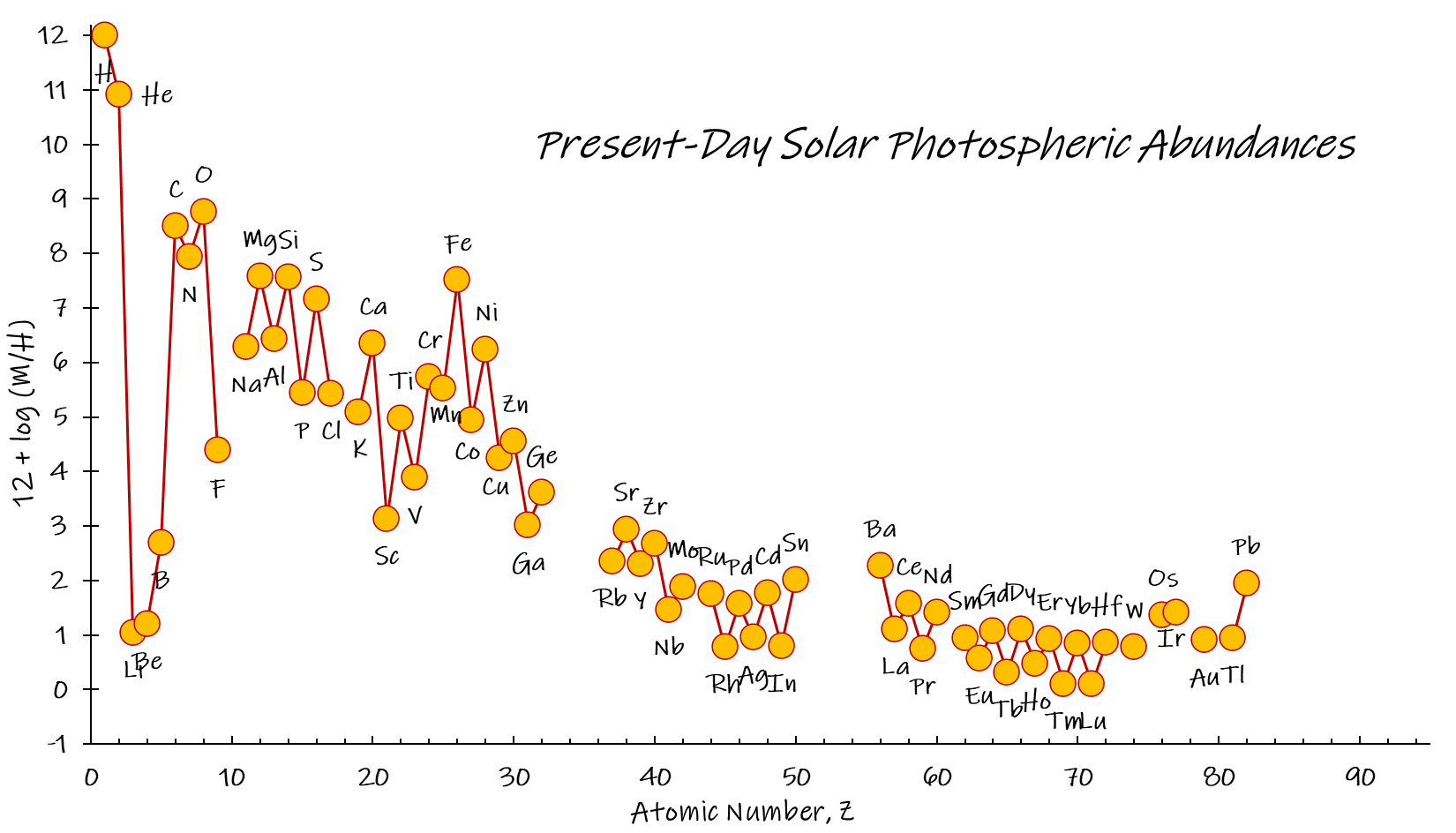}
\caption{Photospheric elemental abundances of the Sun are plotted versus the atomic number of the elements. The abundances are shown on a logarithmic scale where the logarithm of the number of H atoms is set to A(H) $= 12$. The abundances span $12$ orders of magnitude. Not all stable and long-lived (U, Th) elements can be determined from the solar photospheric spectrum.}
\label{chap1:fig1}
\end{figure}

\textbf{Solar abundance} refers to the number of atoms of an element relative to hydrogen in the Sun today. Solar abundances are primarily determined from spectroscopic studies of the solar photosphere, the outer layer of the Sun from which most of its light is emitted. Chemical data of primitive meteorites improve the quality of photospheric abundances. The \textbf{solar system abundances}, or \textbf{proto-solar abundances} are the elemental abundances at the time of solar system formation. 

It is commonly assumed in the astronomical literature that solar photospheric abundances are representative of the bulk solar composition. However, the proto-solar abundances relative to hydrogen, which are the actual abundances of the solar nebula, are somewhat higher than the present-day surface values. This is because the present-day solar surface abundances of heavier elements were modified by gravitational settling over time \citep{Bahcall1992}. The effect is not large and it is assumed that the process does not lead to significant fractionations among heavy elements. But the bulk heavy element composition of the present solar photosphere does not fully represent the initial chemical composition of the solar system nebula, from which the planets and other solid bodies of the solar system formed.

The solar elemental composition can be expressed as mass fractions or as particle fractions. Chemical abundances in the Sun are generally given as ratios that is relative to a pre-defined quantity (by mass or particle number) of a reference element, which means that a reference element is needed. For solar abundances, hydrogen (H) is typically used as reference element, because H is by far the most abundant element in the Sun. The abundance scale is set relative to $10^{12}$ hydrogen atoms, and given on the Briggs’ decimal logarithmic scale (in units of “dex” - dex is a convention to express decimal log scale quantities in astronomy and it is not a unit in physics), the power of 12 is chosen because in the Sun, abundances of a majority of elements vary over 12 orders of magnitude, although some isotopes or measurements in meteorites have negative abundances due to their negligibly small amounts.

Thus, the hydrogen-based \textbf{astronomical abundance scale} for any element $i$ relative to hydrogen is
\begin{align}\label{chap1:eq1}
A(i) = 12 + \log_{10}(N_i/N_{\rm H}),
\end{align}
where N$_i$ is the number density of an element $i$ and N$_{\rm{H}}$ that of H. Commonly the square bracket notation [ ] is used for all other stars,
\begin{align}\label{chap1:eq2}
[i/\rm{H}] = (i/\rm{H})_{\rm{star}} – (i/H)_{\rm Sun},
\end{align}
or preferentially
\begin{align}\label{chap1:eq3}
[i/\rm{Fe}] = (i/\rm{Fe})_{\rm star} – (i/\rm{Fe})_{\rm Sun}.
\end{align}
This applies to any star (see also \citealt{Hinkel2022} for a detailed analysis and derivation). For example, [O/Fe] is the ratio of oxygen to iron in a star relative to the same ratio in the Sun. This ratio represents the total number density of O and Fe particles in all phases, both, atomic and molecular, present in the solar atmosphere. The normalization to Fe is usually used, because the amount of H cannot be directly measured neither in the Sun, nor in any other star. 

The quantification of concentrations requires that the mass or total number density of the analyzed area or volume is known. This number is model dependent. Not all elements can be measured in the solar photosphere with currently available methods. Therefore, the mass fractions of elements for the Sun are calculated by combining evidence from different observational methods and from theoretical solar evolution and interior models (Sects. 3, 4). The mass fractions of H, He, and all other elements (collectively called “metals”), are represented by symbols $X$,$Y$, and $Z$, respectively (see Sect. 3). A fundamental quantity in solar and stellar physics is the metal mass fraction $Z$, which is referred to as solar metallicity (see below). 

Another abundance scale is the \textbf{cosmochemical abundance scale} that is related to $10^{6}$ silicon (Si) atoms. This scale is more practical when objects with low abundances of H and He are investigated. These objects include the Earth, other terrestrial planets, asteroids, comets, and meteorites – typically small objects, which have too low gravity to retain significant amounts of H, He, or which never had high concentrations of volatile gases containing C, N, O or noble gases to begin with, or lost them over time. The cosmochemical scale for an element $i$ is often written as $N(i) = n(i)/n(\rm{Si}) \times 10^{6}$ or sometimes in the astronomical literature as ($i$/Si). Here $n(i)$ is the number of atoms of an element $i$ (and Si) given in parts-per-million (ppm) by number. This is not to be confused with the ppm by mass, which is used for mass concentration measurements in solid matter. 

The astronomical and cosmochemical scales can be converted into each other. The normalization to 10$^{12}$ atoms of H in the astronomical scale is replaced by a normalization to 10$^6$ atoms of Si in the cosmochemical scale. See section 6.1 for the necessary steps to link the solar and meteoritic abundances scales.

The \textbf{uncertainties} for the logarithmic abundance scale are given in dex, so they are, in effect, an uncertainty factor when converted to a linearized scale. This means that an uncertainty on the dex scale typically leads to percentage uncertainties which are not necessarily the same for the higher ($+$) and lower ($-$) bound. The relationship for uncertainties in percent, $U$\%, between the two scales is 
\begin{align}\label{chap1:eq4}
 U(\%) = \pm 100 \times (10^{\pm a} – 1),
\end{align}
where $\pm a$ is the uncertainty in dex on the logarithmic abundances scale, and $U$\% is the uncertainty in percent on the linear scale, e.g., uncertainties of 0.05 and 0.1 dex (here rounded to 2 significant digits) correspond to roughly 10\% and 26\% uncertainties on the linear scale, respectively.

The \textbf{solar metallicity Z} is the combined mass fraction of the heavy elements from Li to U in the periodic table. All elements heavier than He are called “metals” in astronomy and are lumped together because they only make less than 2\% of the total solar mass. The mass fractions of hydrogen and helium are called X and Y, respectively. The total mass of the heavy elements in the entire Sun is the mass of heavy elements the Sun had at its birth $4.567$ billion years ago, and it is little affected by the extent of H to He burning in the Sun’s core. Some small changes occur in the C, N, and O abundances and their isotopic compositions during minor H-burning via the catalytic CN-cycle, but the overall sum of the amount of C, N, and O remains about the same. However, the amount of all the heavy elements observable in the present-day solar photosphere is reduced, because of gravity acting upon heavy elements, as mentioned already above. We come back to this when solar system abundances are discussed below.

The \textbf{relative volatility of the chemical elements} completes the set of basic concepts used throughout this chapter. This definition is based on the distribution of the elements between condensed phases and the gas phase as a function of temperature and total pressure for a given bulk composition. In the astronomical and planetary context, the term volatility is used with respect to the bulk solar composition. Quantitative measures are the 50\% condensation temperatures of the elements that describe the temperature, at which 50\% of a given element is in the gas and 50\% in condensed phases at a given total pressure. \textbf{This quantity is computed by solving the gas-phase and gas-solid thermochemical equilibrium and mass balance for large number of gases and condensates for all chemical elements} \citep[see][]{Palme1990, Lodders2003, Palme2014}. \textbf{Highly volatile} elements only condense at the lowest temperatures ($< 200$ K) and include noble gases, H, C, N, and O, the latter form “ices” such as water H$_{2}$O, methane CH$_{4}$, carbon dioxide CO$_2$, and ammonia. Volatile elements include S, Se, Zn, Cd, and halogens, which condense when S is removed into troilite (FeS) below $\sim 700$ K. \textbf{Moderately volatile} elements, such as the alkali elements and Mn, condense above the troilite formation temperature, but below those of \textbf{common} elements Mg, Si, and Fe \citep[e.g.,][]{Lodders2003}. The latter three elements are the most abundant rock-forming cations, and form Mg-silicates and an iron metal alloy, which make the bulk of condensed “rocky” material. This is why Mg, Si, and Fe are critical, e.g., in studies of planet formation in the solar system and extrasolar systems \citep{Dorn2015}. Oxygen, classified as a highly volatile element is also in the condensing minerals, but oxygen removal by minerals (20-25\%) is limited by the amounts of cations. \textbf{Refractory} elements condense at higher temperatures than the major mineral forming elements; Ca, Ti, Sr, and Eu belong to this group. The highly refractory elements include many transition-elements and the rare-earth elements. Ultra-refractory elements such as Zr, Hf, and Al form the first oxides and W, Re, and Os form the first metallic alloys.
\section{Development of Solar Abundance Studies}

The following discussion describes major advances in converting solar spectra into elemental abundances, transitioning from early qualitative analyses to modern, precise measurements using advanced computational models for 3D modeling of the solar atmosphere and non-local thermodynamic equilibrium radiative transfer (NLTE). Experimental results for atomic and molecular line properties (such as wavelengths, level energies, transition probabilities, damping constants) were equally important to obtain more robust abundances.

Around 1814, \textbf{Joseph von Fraunhofer} discovered characteristic dark lines in the continuous emission spectrum of the Sun, which later have been used to quantify the abundances of the elements in the Sun. He correctly identified these dark lines at certain wavelengths as absorption of light by elements in the cooler outer atmosphere above hotter regions, where the light originated. 
In 1859, \textbf{G. Kirchoff and R. Bunsen} discovered that the lines are element or compound specific and anticipated that the qualitative analysis of the solar spectrum was at hand (Kirchhoff, 1859).

\textbf{Cecilia Payne} (later Payne-Gaposchkin) laid the foundation for the modern understanding of stellar and solar abundances. In her 1925 Radcliffe College thesis “Stellar Atmospheres; A Contribution to the Observational Study of High Temperature in the Reversing Layers of Stars" (Payne, 1925), she established the dominance of hydrogen and helium in the Sun contrary to earlier views that the Sun and stars are similar in composition to Earth. 

\citet{Russell1929} expanded on Payne’s findings and accepted the fact that H and He were the most abundant elements in the Sun, abandoning his earlier statement that the “relative abundance of elements in the universe was like that in Earth's crust” (Russell, 1914). In his 1929 paper, Russell gave the first comprehensive table of solar abundances, but he only briefly acknowledged Payne’s groundbreaking contributions to the quantitative evaluation of solar abundances. 

\citet{Suess1956} presented one of the first studies of elemental abundances in the solar system from solar and meteoritic data. Meteorite data as cosmic abundance standards had already been used by Goldschmidt in the twenties and thirties of the last century (Goldschmidt 1937, 1938), but Goldschmidt did not use solar abundances in his earliest works and thus did not include the most abundant (and volatile) elements H, He, C, N, O, Ne, and Ar. These elements have very low concentrations in meteorites and the solar abundances of C, N, and O can be only obtained from photospheric abundances. Suess and Urey applied nuclear abundance systematics to the abundances of heavier element nuclides that \citet{Suess1947a} and \citet{Suess1947b} had developed in the 1940s, which had been used earlier in rudimentary form by Goldschmidt (around 1920s and 1930s) to constrain several heavy element abundances.

\citet{Cameron1968, Cameron1973} updated elemental abundance tables using constraints from several nucleosynthesis processes to refine abundances of elements that are difficult to measure in the Sun and that cannot be derived from meteorites (e.g., noble gases).

\citet{Anders1989} combined the more precise meteoritic data with spectroscopic measurements of the solar photosphere to create a widely accepted abundance table. 
The solar abundances by \citet{Grevesse1998}, based upon the previous work by this group, were presented in comparison with meteoritic data. Their values provided the baseline composition in the so-called standard solar models, which well described the composition of the solar interior, the energy production, and evolution of the Sun over time. Many comparisons in subsequent works to “older” abundances refer to this abundance set.

The group of \textbf{T. Gehren} and collaborators (\textbf{L. Mashonkina, J. Shi}) set the baseline for quantitative NLTE modelling of solar abundances. Their papers aimed at developing and validating the NLTE models. Careful abundance calculations for individual element provided the first comprehensive NLTE dataset \citep{Gehren1975, Gehren2001, Mashonkina2000, Shi2008, Mashonkina2011}, which has been used as benchmark for establishing the quality of solar NLTE abundance calculations in many subsequent studies.

\citet{Asplund2005} used the 3D Stagger model of the solar atmosphere and NLTE models for several elements, especially C, N, O, and some other abundant elements (e.g., Mg, Si, Fe). Their results suggested that the Sun has a lower metallicity than previously thought. This led to a disagreement of the solar abundances of \citet{Asplund2005} and standard solar models, based on constraints obtained by the solar structure via helioseismology.

\citet{Caffau2011} derived solar abundances of $12$ elements, using independent methods, 3D solar photospheric models computed with the CO5BOLD code, and NLTE atoms for selected species. They obtained higher abundances for the key elements C, N, and O compared to \citet{Asplund2005}. The group of E. Caffau has put substantial efforts in validating the abundances using different atomic and molecular lines \citep[e.g.][]{Ayres2013} and testing the consistency of estimates across the solar disc \citep{Steffen2015}. Standard solar models based on \citet{Caffau2011} composition were explored in \citet{Villante2014} indicating a substantially better agreement with helioseismology and solar neutrino data than earlier models.

\citet{Asplund2009}, \citet{Asplund2021}, \citet{Scott2015a}, \citet{Scott2015b}, and \citet{Grevesse2015} provided further updates on solar abundances with their updated 3D solar model atmosphere and NLTE calculations. Their models often led to lower abundances than recommended by others. 

\citet{Lodders2003}, \citet{Palme2003}, \citet{Lodders2010}, \citet{Lodders2021}, and \citet{Palme2014} updated many meteoritic and solar abundances and their meteoritic values are frequently adopted by other groups.

Over the past decade, the group of \textbf{M. Bergemann} has developed NLTE models and also recently new 3D solar atmosphere models \citep{Eitner2024} in collaboration with the Montpellier, Copenhagen, and Oslo groups of Bertrand Plez, {\AA}ke Nordlund and Mats Carlsson. Some NLTE models by Bergemann and her group (e.g., Cr, Mn, Co, Ba) were used in \citet{Asplund2009} and \citet{Asplund2021}, albeit the choices made in the latter paper partly contradict the recommendations on the use of NLTE models by \citet{Bergemann2019}. As the result of new methodology, improved analyses of solar elemental abundances with more accurate atomic parameters, advanced modeling NLTE effects (scattering, collisions, etc.) in 3D radiation transfer were presented \citep{Bergemann2019, Bergemann2021, Gallagher2020, Magg2022, Storm2024}. 

The latest summary and updated results for many solar elemental abundances are described in detail in \citet{Lodders2025}. The results listed in this paper are adopted for this article.
\section{Solar Composition}

The most abundant chemical elements in the Sun in order of decreasing mass fractions are H (73.9\%), He (24.5\%) and C, N, O, Ne, Fe, which make up 0.28\%, 0.09\%, 0.67\%, 0.21\%, and 0.13\%, respectively. Thus, 7 elements contribute 99.8\% to the Sun’s mass, with small additions from Si, Mg, and S. The contributions of all other elements are negligible. The determination of the element mass fraction from the A(X) values can be done using the following equations, assuming A(H)$=12$ and $i$ designating elements from Li (Z=3) to U(Z=92),
\begin{align}\label{chap1:eq5}
Z/X = \sum (\rm{atwt}(i) × 10^{A(i)}) / (1.00783 \times 10^{12})    
\end{align}
with atomic mass of each element $i$ (e.g. for H, the at. mass $= 1.00783$ Da (Dalton) for the Sun). We note that the Earth has a higher atomic mass for H due to a higher D/H ratio (at. mass $= 1.0079$ Da). With the solar H mass fraction of $X = 0.7389$ and the solar photospheric abundances A($i$) from \citet{Lodders2025}, we obtain a $Z/X = 0.02162$ and a total solar metallicity of $Z = 0.0160$ or 1.60\%. 

Finally, $Y$, the total mass fraction of He, can then be derived from the standard equation as $Y = 1 - X - Z = 1 - 0.7389 - 0.0160 = 0.2451$. We come back to these mass fractions in section 6.2.

The elements heavier than helium are called “metals” in astronomy but the astronomer’s term “metals” is unfortunate, because not all elements heavier than helium are metals in the conventional sense (i.e., good conductors of electricity and heat). For example, the halogens (F, Cl, Br, I) are not usually regarded as metals in chemistry and physics. Also among pnictogens (N, P, As, Sb, Bi), N and P are not metals, nor are O and S metals. The latter two belong to the group of chalcogens (O, S, Se, Te). In chemistry and physics, P, Se, and Te are regarded as metalloids or p-block elements. 

The solar abundances are still associated with comparatively large uncertainties and only about a quarter of the elements measured in the photosphere are determined with full 3D radiation transfer models and corrected for NLTE effects (see section 5). 
\subsection{Abundances in the Photosphere}

Table \ref{tab:1} lists photospheric elemental abundances on the atomic astronomical and cosmochemical abundance scales. For comparison, data for CI-chondrites discussed below in section 5 are also listed here on the same abundance scales.

Figure \ref{chap1:fig1} shows the photospheric abundances of elements that can be determined quantitatively. Hydrogen plots at 12 on the decadic logarithm scale. The helium abundance is not a photospheric value but instead is derived from helioseismology (see below). The He value is shown here to demonstrate the large He abundance in comparison to all other heavy elements. There are several “gaps” in the curve in Figure \ref{chap1:fig1} notably for noble gases other than helium. The noble gas values have to be estimated by indirect means using solar wind data or nuclear systematics (see below). For other “missing” values of stable elements, and for the long-lived element U and Th meteoritic values are used to obtain the full set of abundances. 

Figure \ref{chap1:fig1} also shows that atoms with even atomic numbers are more abundant than atoms with odd atomic numbers. The even-odd effect is clearly seen in the figure and it is well-established for many Galactic stars. This effect demonstrates that the abundances of elements in the Sun, in stars, and in CI chondritic meteorites are the result of stability and structure of nuclei. This has important consequences for stellar nucleosynthesis. Thus, the even-odd effect is related to nuclear stability which in turn controls nucleosynthesis yields (see section 6.3).
\section{Methods to Determine Solar Abundances}

\begin{figure}[t]
\centering
\hbox{
\includegraphics[width=0.5\textwidth]{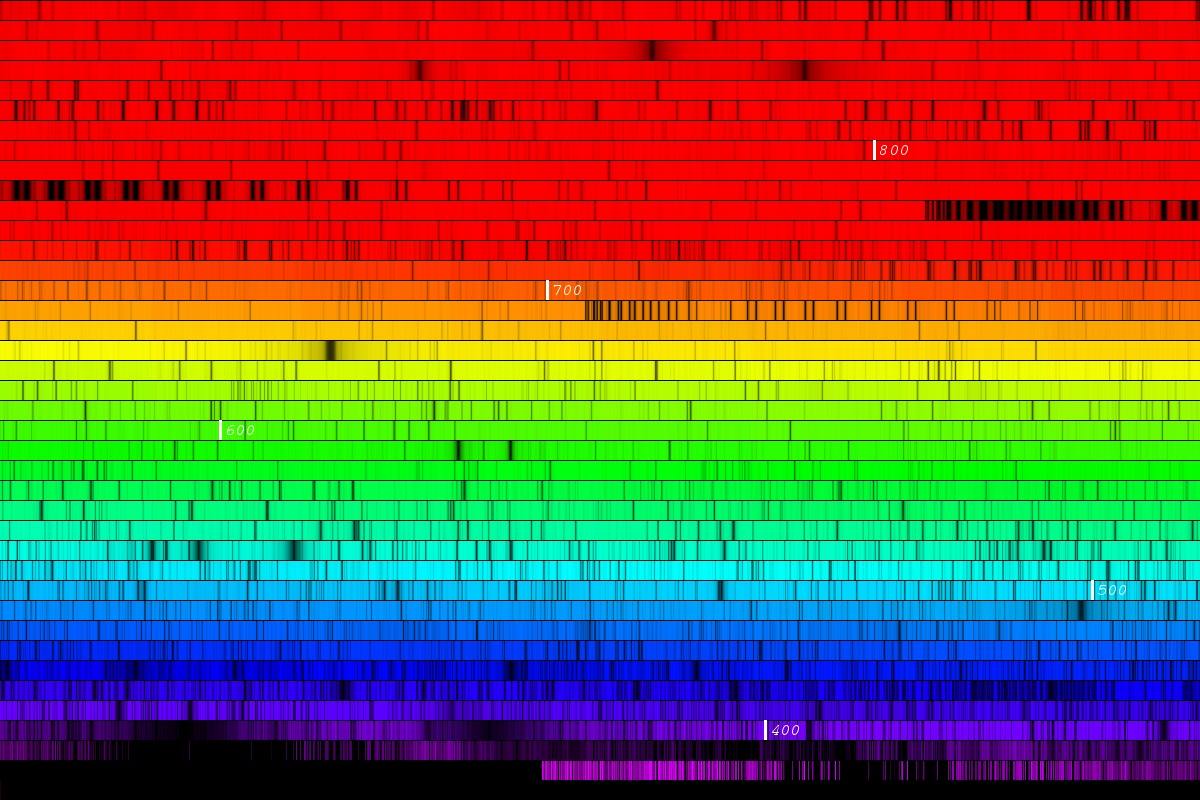}
\includegraphics[width=0.5\textwidth]{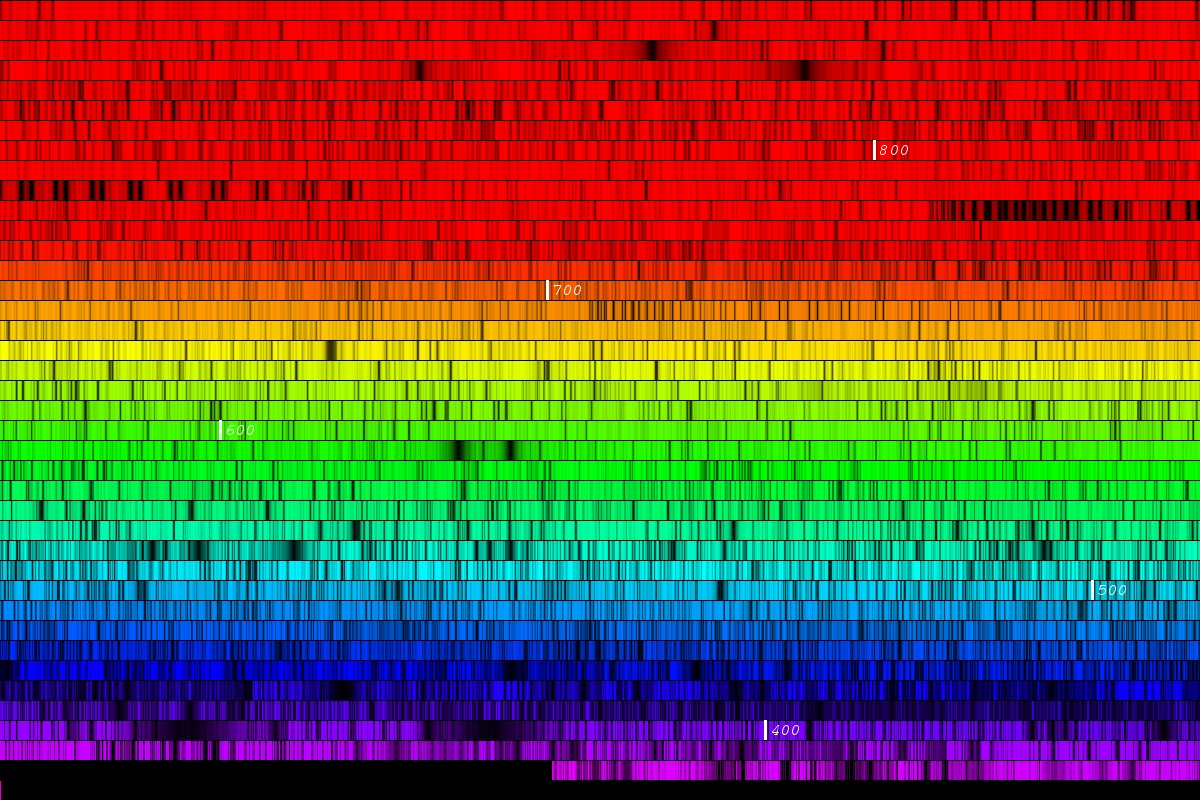}}
\caption{Observed spectra of the Sun (left) and a metal-rich red giant $\mu$~Leo (right). The data were obtained with the NARVAL spectrograph at Pic du Midi \citep{Blanco-Cuaresma2014}. The stellar parameters of $\mu$~Leo are $T_{\rm eff} = 4474 \pm 60$ K, $\log g = 2.51 \pm 0.09$ dex, and [Fe/H] $=0.25 \pm 0.15$ dex \citep[for details, see ][]{Jofre2015}. The ordering is from the near-UV (bottom right) to near-IR (top left), and vertical white marks indicate wavelengths in nano-meters. Such spectra are typically used for the physical analysis of solar and stellar chemical abundances. Figure (c) M. Bergemann and S. Brinkmann, MPIA.}
\label{chap1:fig2ab}
\end{figure}

Elemental abundances in the Sun's photosphere are mainly determined from absorption lines, such as the Fraunhofer lines in the solar spectrum, measured against the continuum radiation (background light emitted from the Sun’s surface). The solar photospheric spectrum – the electromagnetic radiation emitted from the solar photosphere as a function of wavelength – ranges from ultraviolet (UV, $\sim$ 300 nm) to infrared (IR, $\sim$ 3000 nm $= 3$ microns). Similar spectroscopic data are now available for many Galactic stars. Figure  \ref{chap1:fig2ab} shows the high-resolution spectrum of the Sun and a red giant $\mu$ Leo; both spectra taken with the NARVAL facility.

Abundances of chemical elements in the Sun have mainly been derived from solar spectra at visible wavelengths, although for some very heavy elements also the UV solar spectra were used. For selected molecular lines (e.g., di-atomic species like OH or CO, and their isotopologues) the infra-red data are sometimes used \citep{Hall1972, Ayres2013}. Each element absorbs light at characteristic wavelengths, and the depth and shape of these absorption lines can be used to infer the abundance of each element. Currently absorption lines of around 60 elements can be detected, isolated, and quantified in the photospheric spectrum. Spectra from cooler sunspots provide abundance estimates for F, Cl, In, and Tl. 

For all chemical abundance analyses of the solar spectrum, which is the method of choice in all solar compilations \citep{Grevesse1998, Asplund2009, Caffau2011, Magg2022}, assumptions for the physical structure of the solar atmosphere have to be made. Spatially-dependent distributions of temperature, pressure, density, velocities, and break-down of individual species into chemical compounds, and the atomic properties of the elements need to be known in order to compute the solar synthetic spectra. Different physical models are used (Section 5). The solar photospheric elemental abundances are then found by varying the input elemental abundances until a good match to the observed spectrum is found.
\subsection{Spectroscopic Techniques and Instrumentation}

Different techniques and instruments are used to measure the solar radiation. 

Most of the photospheric abundance measurements rely on wide-band high-resolution solar spectra obtained with ground-based facilities, including (but not limited to) the Kitt Peak National Observatory (KPNO) Fourier Transform Spectrograph (FTS, Kurucz et al. 1984) and recently also the solar atlas obtained via FTS observations with the telescope of the Institute of Astrophysics G\"{o}ttingen \citep[IAG, ][]{Reiners2016}. These atlases contain data of the solar spectra either as bulk fluxes (integrated over the entire solar disc) or as spatially-resolved intensities for different positions (angles) across the solar disc. Therefore, abundance measurements can be carried out for different pointings separately and consistency between measurements serves as an important quality control and validation for solar chemical abundances \citep{Steffen2015, Lind2017, Bergemann2021}.

However, ground-based spectral instruments are of limited use in certain wavelength ranges, because of interference with absorbers (water, CO$_{2}$) and scattering by molecules and aerosols in the Earth’s atmosphere. Even when instruments are located at high altitudes, these interferences are still present. The high-energy portions (UV) of the solar spectrum is absorbed by Earth's atmosphere and this wavelength range is only accessible through rocket-bound and space-bound instruments. 

Space-based measurements avoid atmospheric distortion and allow precise solar spectrum measurements. Instruments dedicated to observations of the Sun are the Solar and Heliospheric Observatory (SOHO), the Solar Dynamics Observatory (SDO), the European Space Agency's (ESA) Solar Orbiter and ultraviolet and Extreme Ultraviolet Spectrometers. 

One future instrument for studying the Sun is the 4.2-m European Solar Telescope (EST) in the Canary Islands.

Spectrometers and spectrographs split sunlight into its component wavelengths using prisms or diffraction gratings. The dispersed light is then detected by sensors, e.g., photodiodes, charge-coupled devices (CCDs). Photometers use specific filters to measure the intensity of sunlight in narrow wavelength bands. 

Direct measurements of the solar irradiance utilize pyranometers and spectroradiometers, whereas indirect measurements use models or algorithms to reconstruct the spectrum based on atmospheric and satellite data.

The instruments are calibrated and standardized against standard light sources to ensure accuracy. Solar spectrum measurements are often reported in terms of “solar irradiance” in units of W/m$^2$/nm, which gives the energy radiated per second (J/s $=$ W) per square meter (m$^2$) and per wavelength (nm). In addition to deriving elemental abundances from the solar spectrum, measurements of the solar radiation are increasingly important for solar energy technology design and monitoring space weather.

Observations of asteroids (e.g. Ceres, Vesta) or moons of solar system planets (e.g., Ganymede) are sometimes used as independent probes of the solar chemical composition. These bodies provide reflected spectra of the Sun, thus acting as ‘mirrors’. Primarily, however, they are useful in order to place stellar abundance measurements onto the same chemical abundance scale as the Sun. Asteroids are much fainter than the Sun, and therefore can be observed with the same spectroscopic facilities as those typically used for observations of stars in the Milky Way galaxy \citep{Blanco-Cuaresma2014, Jofre2015}. These facilities include HARPS at ESOs 3.6m telescope in La Silla, UVES and ESPRESSO at the Very Large Telescope, and PEPSI at Large Binocular Telescope \citep{Adibekyan2020}.
\subsection{Helioseismology}

The H and He content of the Sun, as well as the metallicity of the solar interior are  probed by helioseismology, which is the method to study solar oscillations (see also the Chapter on Helioseismology in this Encyclopedia). 
Helioseismic models describe the Sun's internal structure from measured sound speed profiles, which place constraints on the Sun’s internal composition and how the Sun evolved over time to its current state. The composition is important because the heavy elements determine much of the opacity in the solar interior. These models also allow the derivation of the solar helium abundance, which is a free parameter in the models \citep{Basu2004, Basu2008}. Since the early 2000s, there have been debates about the discrepancy between the solar composition derived from helioseismology and that from spectroscopic observations. This discrepancy stimulated much new work on abundance measurements, new measurements and revisions of atomic data, and improvements of models on the solar interior. 

\citet{Basu2004} and \citet{Basu2008} investigated the impact of the revised, lower solar abundances of C, N, O, and Ne on helioseismic models, and discussed the conflicts between observed and modeled solar interior properties using lower abundances of C, N, O, and Ne, because these do not provide enough opacity in the Standard Solar Model (SSM) to satisfy the observations. Currently the application of helioseismology to derive solar abundances of heavy elements is still advancing.

For the key chemical elements, the solar photospheric abundances from \citet{Bergemann2021}, \citet{Magg2022}, and \citet{Lodders2025} are higher compared to the data presented by \citet{Asplund2009} and \citet{Asplund2021} and thus closer to pre-2000 abundances, which worked well for the standard solar models. Thus, the conflict between helioseismic data and solar models may be resolved. Especially in the sound speed profile there is now a much-improved agreement between helioseismic results and the SSM models based on the new composition from different stellar-evolution codes such as YREC and GARSTEC \citep[e.g.][]{Magg2022, Yang2024, Basinger2024}. Gravitational settling efficiency of metals from the solar photosphere is still a source of uncertainty. The analysis of the first adiabatic exponent profile of the solar interior \citep{Baturin2024} is inconclusive. It lends some support to lower C and O abundances in agreement with \citet{Asplund2021}, and for O also with \citet{Bergemann2021} within the uncertainties of both results. However, the analysis by \citet{Baturin2024} also favors significantly high solar N and Ne abundances in agreement with \citet{Magg2022}.
\subsection{Solar Neutrinos}

Neutrinos are produced during H fusion to helium via the CNO cycle in the solar core. Although this H fusion mechanism is not the major process producing neutrinos (proton-chain (pp$-$) fusion being the major mechanism), the catalytic reactions involving C, N, and O release neutrinos in quantities that are proportional to the quantities of C, N, and O present. As already mentioned, C, N, and O are the major “metals” in the Sun and their abundances determine much of the solar opacity.

\begin{figure}[t]
\centering
\includegraphics[width=1\textwidth]{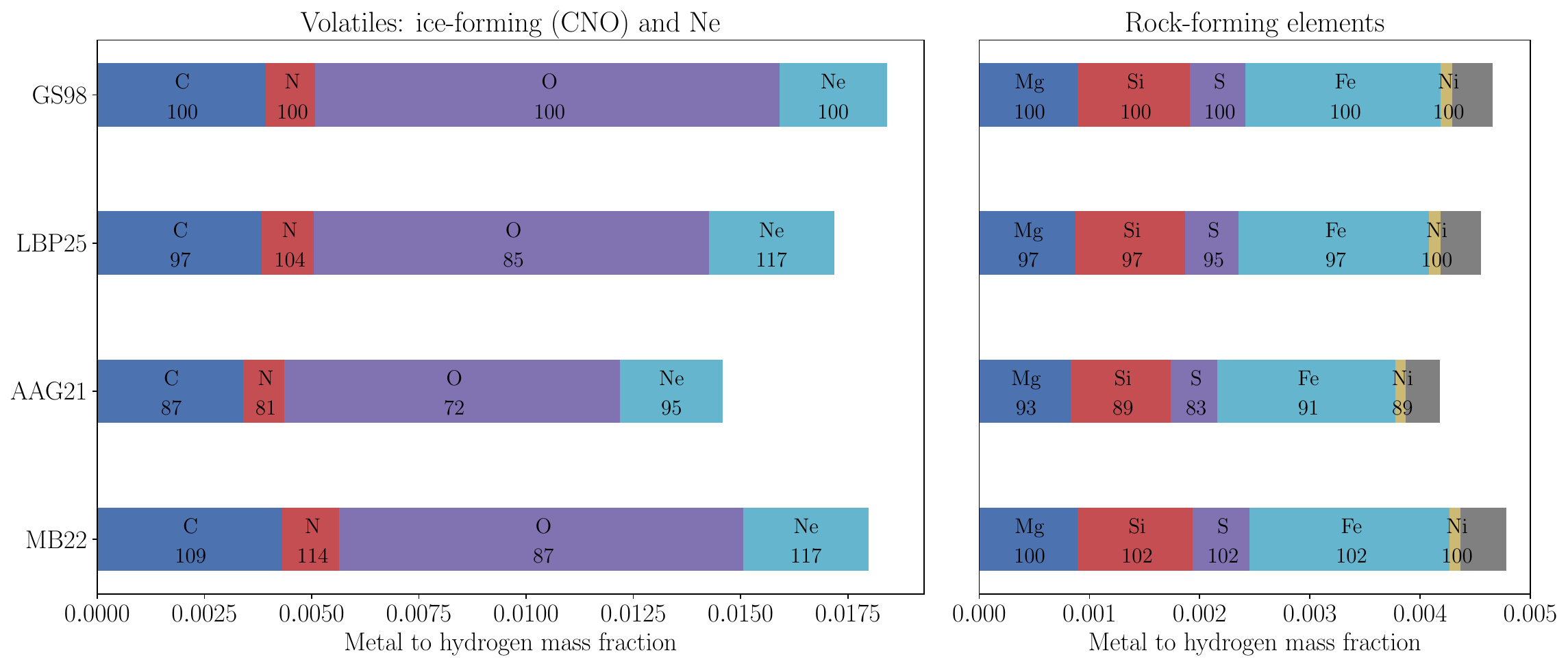}
\caption{Metal to H mass fractions for key chemical species, including volatiles (C, N, O, Ne) and some non-volatiles (Mg, Si, S, Fe, Ni), based on solar photospheric measurements by different authors: GS98 (Grevesse \& Sauval 1998), AAG21 (Asplund et al. 2021), MB22 (Magg et al. 2022), and LBP25 (Lodders et al. 2025). The x-axis gives the mass fraction of individual elements, as indicated in the coloured regions. The numbers in the coloured regions represent the percentages relative to $A(i)$ values from Grevesse \& Sauval (1998). Hence on the top panel each element is indicated as 100\%. The compilations by Asplund et al. (2021), Magg et al. (2022), and Lodders et al. (2025) have lower O mass fraction (by 28\%, 13\%, and 15 \%, respectively) than GS98. Figure courtesy A. Serenelli}
\label{chap1:fig2}
\end{figure}

\textbf{J.N. Bahcall and collaborators} (see also the Chapter 2 on this topic in this Encyclopedia) explored the connection between solar abundances and solar neutrino fluxes. They discuss the discrepancies between solar models done with lower solar abundances and neutrino observations \citep{Bahcall1982, Bahcall2006, Serenelli2004, Serenelli2010}. Recent neutrino fluxes resulting from the pp-chain and CNO cycles in the Sun serve as advanced testable observables, which were reported by \citet{Appel2022} and \citet{Basilico2023}. Different neutrino fluxes have been measured with Borexino particle experiment, including pp, pep, $^{7}$Be, and $^{8}$B, the latter two very accurately to 3.5\% and 2\%, respectively, allowing stringent constraints on the structure of the solar interior. The combined C$+$N abundances derived from the BOREXINO neutrino detector experiment are $5.81^{+1.22}_{-0.94} \times$ 10$^{-4}$ \citep{Basilico2023}. They note that among various photospheric abundances, the best agreement is to the photospheric (C$+$N)/H ratio by \citet{Magg2022}. For the visual illustration, we show in Figure \ref{chap1:fig2} the comparison of mass fractions for the key elements taken from three different compilations. 

Our recommended C abundance from \citet{Lodders2025} is somewhat lower than in \citet{Magg2022}. Our photospheric (C$+$N)/H ratio of $= 4.11(\pm 0.7) 10^{-4}$ also agrees with the Borexino values within error limits. In order to compare both values, one has to correct the present-day photospheric abundances for gravitational settling. The correction is roughly $23 \%$ and it is described in Sect. 8.2. The solar interior essentially remained at the protosolar values for C and N, and the amount gained in the interior from settling is very small in absolute terms, because the convective envelope is only about 2\% of solar mass. Thus, we compare the BOREXINO results to our protosolar ratio (C$+$N)/H $= 5.03 (\pm 0.97)10^{-4}$ (Table \ref{tab:4}). This is closer to the nominal BOREXINO value (within 13\%) and also indicates that the larger settling corrections could be plausible. 
\subsection{Solar Corona and Solar Wind}

\begin{figure}[t]
\centering
\includegraphics[width=0.8\textwidth]{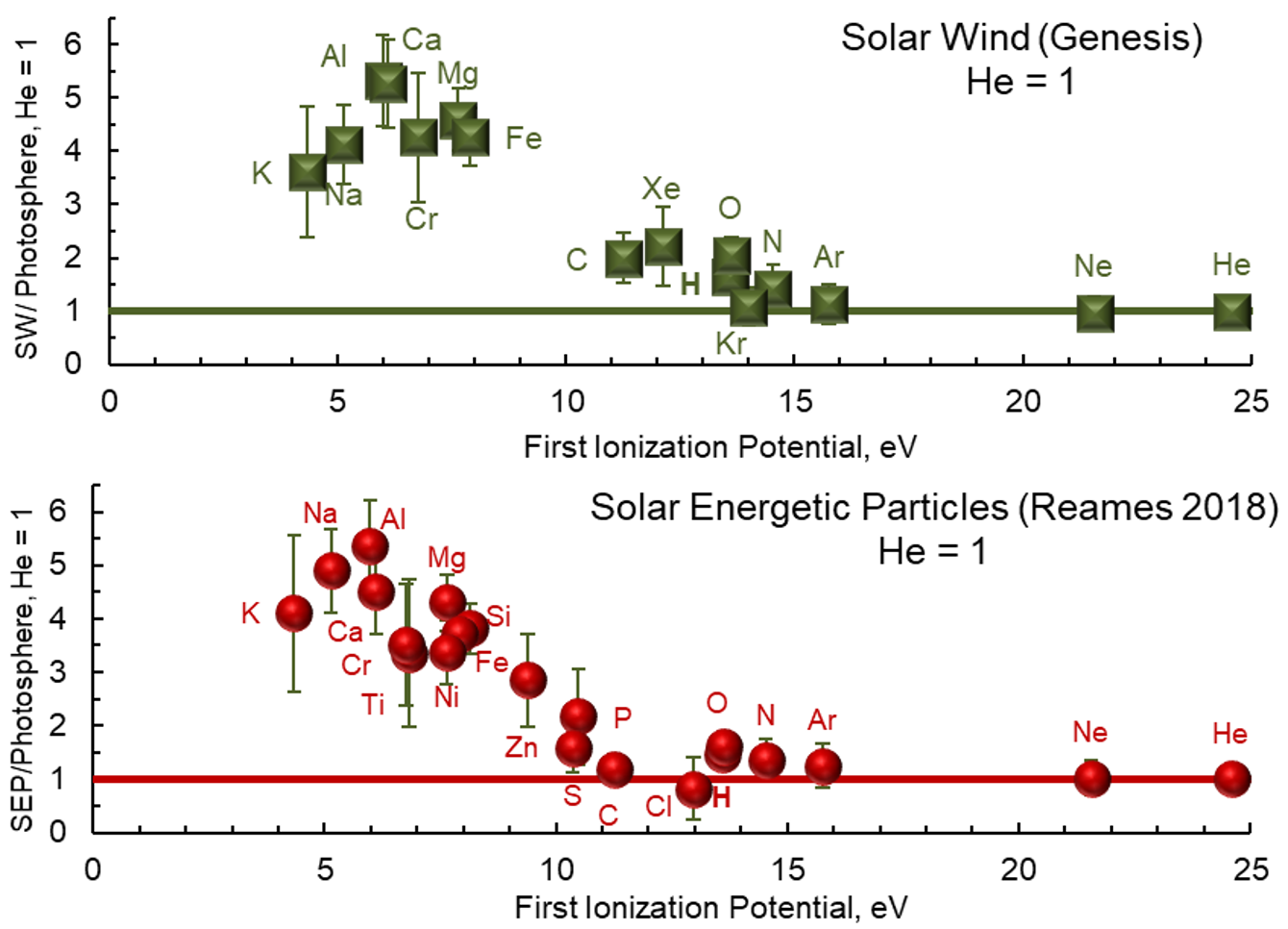}
\caption{Solar wind from Genesis \citep[][and references therein]{Heber2021, Jurewicz2024}, and solar energetic particle \citep{Reames2018} abundances normalized to solar photospheric abundances and helium as a function of first ionization potential (FIP) of the elements. Overall the SW and SEP abundances measured in different ways look very similar. Easily ionizable atoms appear enriched in this normalization. If the same data were normalized to an element with low FIP like Mg or Fe, the low FIP elements would plot at unity and look “normal” whereas the high-FIP elements would appear depleted.}
\label{chap1:fig3}
\end{figure}

The solar corona and solar wind are components of the Sun’s outer layers and are sourced from the photosphere. Accurate measurements of their composition are still challenging. The Genesis mission returned solar wind collected in aerogel collectors in space and results are limited to mainly abundant elements H, Mg, Fe, Cr, Ca, Al, Na, K, and the noble gases \citep[][and references therein]{Huss2020, Meshik2020, Heber2021, Jurewicz2024}. The Genesis results for the bulk solar wind are similar to the abundances of solar energetic particles (SEPs) as described in \citet{Reames2018}. A comparison of these data relative to photospheric values is shown in Figure \ref{chap1:fig3}. Both, solar wind and SEP abundance data are given in Table \ref{tab:2}. Ongoing missions like the Solar Orbiter (SoLO) and the Parker Solar Probe (PSP) collect detailed data on the solar wind and coronal composition and will improve our understanding of how the solar wind carries material from the Sun into the solar system.

The SEPs are high-energy charged particles emitted by the Sun during solar flares and coronal mass ejections. Like the solar wind particles measured by Genesis, the SEPs consist of mainly protons, electrons, and ions of all other heavy elements with similar orders of magnitude in abundances as found in the photosphere. But there are  also important differences in abundances when compared to the photospheric abundances. During their removal from the photosphere and acceleration the coronal and solar wind abundances become fractionated relative to the photospheric composition. These fractionations correlate with the first ionization potential of the elements and indicate that hydromagnetic processes are involved. Enrichments of elements with low FIP can arise from Alfenic waves causing ponderomotive forces (forces in spatially inhomogeneous, electromagnetic fields acting on moving charged particles) that accelerate chromospheric ions \citep[see e.g.,][]{Dahlburg2016,Laming2015, Laming2017a, Reames2018}. Thus, model-dependent adjustments are necessary to correct the solar wind abundances back to photospheric values for elements such as the noble gases that cannot directly be measured in the photosphere.

\section{Advances in Spectroscopic Analysis: 3D Models and Non-LTE Effects}

Traditional methods for determining solar abundances relied on so-called one-dimensional (1D) stellar atmosphere models. These are plane-parallel physical models of stellar atmospheres, calculated using the known effective temperature, surface gravity and detailed chemical abundances of the Sun as input, along with opacities calculated using some free parameters. These 1D models have served as the basis of thousands of studies in astronomy and provided a flexible and versatile method to calculate chemical composition of all kinds of stars. However, 1D models are physically highly simplified. They are usually semi-infinite 1D strings stretched horizontally (in non-radial direction), in order to calculate radiation transfer at different angles with respect to the radial direction and they assume different types of physical and chemical equilibria, including hydrostatic equilibrium, chemical equilibrium, and importantly local thermodynamic equilibrium (LTE). Complex physical processes, like convection, magnetic fields, pulsations or oscillations, and non-equilibrium processes cannot be included in such models from first principles. Convective energy transport is parameterized via the Mixing-Length Theory \citep{Boehm-Vitense1958}, the cumulative effect of turbulence on opacity is described by the free parameter of ‘micro-turbulence’, whereas the effect of turbulence on overall line intensities is parameterized by the free parameter ‘macro-turbulence’. These free parameters are partly defined such that they describe some structural quantities of the Sun \citep{Fuhrmann1993, Bernkopf1998, Gustafsson2008}, but are not adequate to describe consistently the entire electromagnetic radiation of the Sun, including its center-to-limb variation, surface granulation and its variability, detailed physical structure of the atmospheres, and individual line profiles, which are central to solar abundance determinations \citep[e.g.][]{ Dravins1982, Asplund2005, Nordlund2009, Bergemann2012, Bergemann2021, Lind2017, Steffen2015}. 
Major improvements over the past decades are the integration of 3D radiation-hydrodynamic models of stellar atmospheres and the inclusion of non-local thermodynamic equilibrium (non-LTE) radiation transfer in the context of physically-realistic modelling of the solar atmosphere.
\subsection{3D Radiative Transfer Models}

3D radiative transfer models aim to simulate how light is absorbed and emitted across the Sun’s atmosphere by including more realistic representations of gas dynamics and physical variations of temperature, density, and pressure, and even magnetism \citep{Nordlund2009, Carlsson2016, Leenaarts2020}. The coupled solutions of time-dependent radiation-hydrodynamics equations in 3D yield detailed physical structure of the solar subsurface layers and its (very limited due to major computational overheads) time evolution. The upper part of these numerical simulations, typically of several mega-meters in size, covers the entire solar photosphere. When magneto-hydrodynamics (MHD) is used, such models extend to greater outer scales and also include the chromosphere and corona. The physical structure is then used to calculate detailed model spectra of the Sun, including more comprehensive treatment of multi-wavelength opacities, thereby yielding much more realistic and accurate abundances for elements like O, C, N and other species.

The formation of lines in 3D radiation-hydrodynamics (RHD) solar or stellar models is typically such that the line strengths become stronger, compared to line strengths predicted by 1D hydrostatic models. This is because the temperature and density distribution of 3D RHD models is set by hydrodynamics and a delicate balance between radiative heating and cooling, the latter set by opacities \citep{PerdomoGarcia2024}. These lead to the sustained presence of warmer granules but also colder intergranular lanes. In the intergranular lanes, temperatures are lower by 1000s of K, compared to the average 1D hydrostatic structures, which are for the solar metallicity nearly in radiative equilibrium. As a consequence, 3D RHD models are characterized by larger molecular number densities \citep{Uitenbroek2011}, as lower gas temperatures and higher densities allow for a more efficient formation of molecules in the outer photospheric layers. Large concentrations of molecules imply stronger absorption lines in models, and hence smaller abundances of species obtained from the solar spectral analysis, although the effect depends on molecular transitions under consideration. This effect is also known for abundances based on atomic lines.

The calculations of abundances based on molecular diagnostic lines \citep{Asplund2004, Asplund2005b} yielded systematically lower abundances of elements like O and C. However, these estimates were based on the strict assumption of LTE, despite early cautionary notes against this approach \citep{Eugene-Praderie1960}. LTE assumption in molecular line formation was deemed inadequate once first detailed NLTE calculations for molecules were carried out \citep{Popa2023}. 

As standalone, 3D radiation transfer calculations in LTE are not sufficient to provide realistic abundances. This is evident, among other problems, by the huge systematic error of over $0.5$ dex in abundance that 3D LTE modelling implies for the disc variation of the solar intensities and hence fluxes \citep{Bergemann2021, Pietrow2023}. For vast majority of elements, it has been demonstrated in many detailed theoretical and observational studies of the Sun that only NLTE radiation transfer, ideally coupled with 3D structure models, can provide accurate solar chemical composition \citep{Kiselman1993, Kiselman1995, Asplund2005a, Caffau2009, Amarsi2017, Lind2017, Bergemann2019}.

\subsection{Non-LTE Effects}

NLTE calculations of radiation transfer are essential to derive accurate solar abundances. The bare physics of NLTE is the same one as that responsible for the fact that stars emit any electromagnetic radiation \citep{Mihalas1973} and hence are visible to our eyes at all. 

Radiation field interacts with all gas particles in the solar atmosphere via photo-excitation and photo-ionization processes and their reverse. This electromagnetic interaction changes the internal distribution of species among their energy micro-states, that is, via quantum mechanics, the distribution of particles in terms of their principal quantum numbers, orbital angular momenta, and spin values. For molecules, also photo-induced dissociations and attachment processes trigger further departures from LTE, via impact on distributions of number densities of states of different energies, and thus of their electron and orbital angular momenta, but also on distributions regarding the symmetry of micro-states, and rotational and vibrational quantum numbers.

NLTE radiation transfer calculations can be carried out in any geometry, also in 3D using the 3D radiation-hydrodynamics simulations of the solar atmosphere. NLTE effects influence the strength and shape of spectral lines, especially for elements like O, Na, Mg, Al, K, Ca, Mn, Fe, Ni. Therefore, neglecting NLTE may lead to significant errors in solar abundance determinations. The amplitude and sign of NLTE effect on abundances in the solar atmosphere might change drastically depending on the properties of energy states involved in the transition. For some lines, like the main diagnostic lines of O I around 777 nm, the LTE assumption massively over-estimates the abundance by over 50\%, that is 0.2 to 0.3 dex \citep{Bergemann2021}. For the same element, the forbidden [O I] line is, however, almost insensitive to NLTE \citep{AllendePrieto2001, Caffau2008}. Similar differential effects are known for other species, like Li \citep{Carlsson1994}. Many atomic species experience substantial over-ionization due to low ionization energy and large ionization cross-sections in the optical and near-UV. Such species, e.g., Na I, Mg I, Ca I, Cr I and Fe I, tend to be present in much smaller concentrations when NLTE effects are taken into account. Hence their LTE abundances are usually under-estimated \citep{Korn2003, Mashonkina2007, Mashonkina2011, Bergemann2010}. NLTE effects have been explored in many detailed studies, and we refer the reader to the reviews by \citet{Asplund2005}, \citet{Bergemann2014}, and \citet{Lind2024}. Examples for non-LTE effects on Fe I and Fe II lines in the Sun and metal-poor stars are provided in \citet{Bergemann2012} and for Mn in \citet{Bergemann2019}; the latter refining the solar Mn abundance, which had been a puzzle for a long time when compared to meteoritic abundances.

Thus, traditional LTE-based solar abundances may often be too low. Overall, the use of Non-Local Thermodynamic Equilibrium (NLTE) radiation transfer improves the accuracy and precision of spectroscopic abundance determinations, especially for trace elements, but such calculations are computationally extensive. The use of 3D RHD models jointly with NLTE further improves abundance accuracy, but in many cases can be avoided through a careful analytical approach.
\section{Meteorites and the Significance of CI-Chondrites for Deriving the Composition of the Sun}

\begin{figure}[t]
\centering
\includegraphics[width=0.6\textwidth]{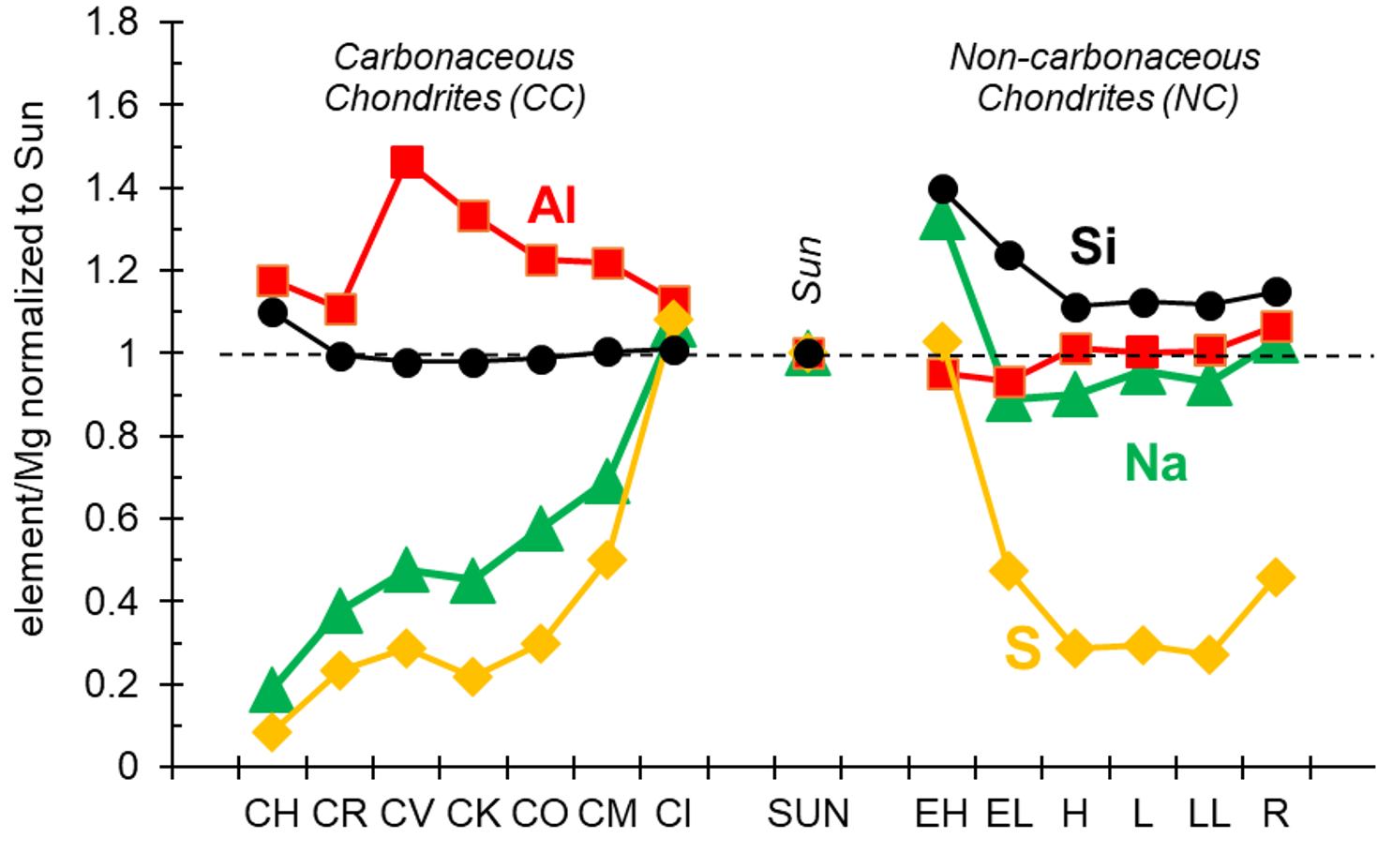}
\caption{Comparison of elemental abundances in different chondrite groups and the Sun. Shown are element/Mg ratios relative to the respective solar values. The Sun plots at unity. Each of the four elements shown is representative of groups of elements with similar volatility (see text). The CI-chondrites show the best agreement to solar values. Other types of meteorites are described in \citet{Krot2014}.}
\label{chap1:fig4}
\end{figure}

There are two types of meteorites, differentiated and undifferentiated ones. Differentiated meteorites are from once-melted planetesimals. Undifferentiated meteorites, such as chondrites, never were heated to (full) melting temperatures and represent aggregation of primary solar system material. The comparatively uniform composition of chondrites should record the average composition of the Solar System (for the classification of chondritic meteorites, see \citealt{Krot2014}). The quest for the representative chondrite group has a fairly long story, and the best choice is CI-chondrites \citep{Lodders2021}. This choice was largely dictated by the differences in abundances among chondrite groups which mainly concern volatile elements such as e.g., Na, Mn, and S (see Fig. \ref{chap1:fig4}).

In Figure \ref{chap1:fig4}, the compositions of different chondrite groups are compared to solar photospheric abundances from \citet{Lodders2025}. The four elements/Mg ratios are selected to represent groups of elements with similar volatility. Aluminium stands for refractory elements such as Ca, Sc, Ti, and most REE. Silicon has a similar volatility as Mg, and Fe. Na and S stand for moderately volatile elements such as other alkali elements and Mn. The four element/Mg ratios are further normalized to the solar respective ratios and plotted for each group among the carbonaceous chondrites and the non-carbonaceous chondrites. The Sun plots at unity and is shown in the center of the diagram. Any deviations from unity indicate enrichments or depletions in chondrites, e.g., Al/Mg $>$ 1 is common in most carbonaceous chondrites. Depletions of volatile elements, e.g., S/Mg $<$ 1, are observed in most chondrite groups. 

Deviations from solar, if any, are smallest for carbonaceous chondrites of the “Ivuna type” or “CI-chondrites”. For example, their Na/Mg and S/Mg ratios are identical within uncertainties to the respective ratios in the Sun. Many other element/Mg ratios in CI-chondrites show excellent agreement with photospheric abundances (see below) which is not the case for other chondrite groups. This is the reason why CI-chondrites are singled out and are used as a solar system abundance standard. A comprehensive discussion of CI-chondrites is given in \citet{Lodders2025} and references therein. 
\section{Solar System Abundances from Combined Solar and Meteoritic Data}

As seen in Table \ref{tab:1}, not all elements can be quantitatively measured in the solar photosphere or in sunspots. Therefore, solar spectroscopic and meteoritic data are combined to obtain a complete set of the stable and long-lived elements of the periodic table, representing the average solar system composition.

\begin{figure}[t]
\centering
\includegraphics[width=0.8\textwidth]{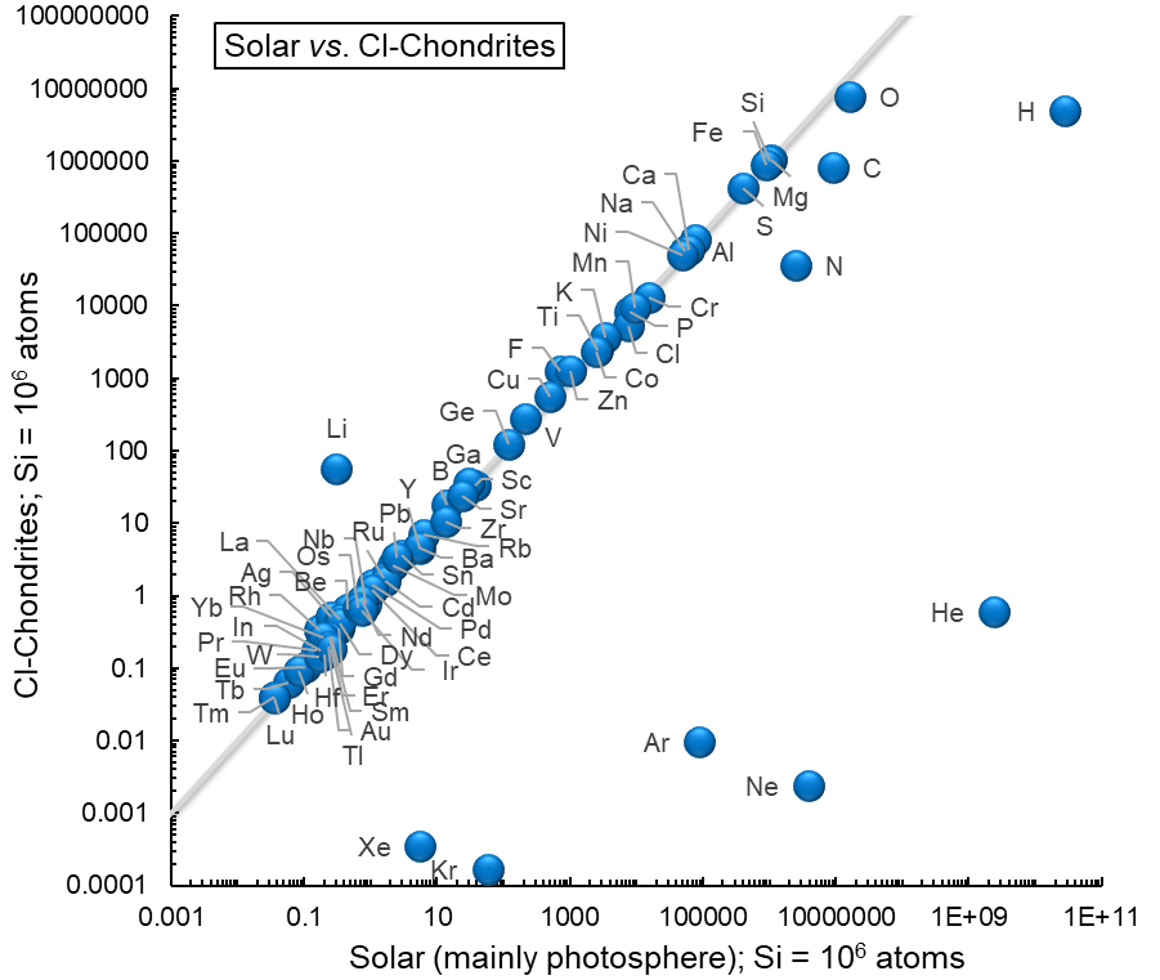}
\caption{Comparison of solar and CI-chondritic elemental abundances of elements where the comparison can be made. CI-chondrites are excellent proxies for the solar composition, except for the most volatile elements H, C, N, O, and the noble gases, and for Li, which is destroyed in the Sun. The abundances are taken from Lodders et al. (2025).}
\label{chap1:fig5}
\end{figure}

It has been known for some time that the elemental abundances in the rare group of CI-chondrites match best with solar photospheric abundances where the comparison is made. This is shown in Figure \ref{chap1:fig5}, where solar and meteoritic abundances are plotted against each other. The comparison spans abundances over 12 orders of magnitude, thus the agreement on these scales is impressive. Overall, the relative concentrations for several well-determined elements in CI-chondrites and in the photosphere are very similar, which is the major argument that CI-chondrites are a good proxy for the condensable elemental abundances, including those that currently cannot be determined in the Sun. The obvious outliers are elements that are highly volatile (noble gases) and elements that form highly volatile gases (C, N, O); these elements are not (fully) retained in meteorite parent bodies. Another exception in Figure \ref{chap1:fig5} is Li, which is about 190-times lower in the solar photosphere than in chondrites. This is due to pre-main sequence Li destruction and ongoing settling combined with further nuclear destruction of the fragile lithium nuclei at the hot bottom of the solar convection zone. Similarly, Be and B could be affected by this type of destruction and mixing \citep{Boesgaard2005, Boesgaard2016}, but within the large uncertainties, the CI-chondritic/solar ratios for Be $= 1.0 \pm 0.4$ and B $= 1.3 \pm 1$ suggest no substantial Be and B depletion in the solar photosphere.

Recent results from the sample return missions from asteroid Ryugu (Hayabusa 2 mission) and asteroid Bennu (OSIRIS-REx mission) showed that these asteroids are compositionally very close to CI-chondrites. These results may improve the "meteoritic" solar composition in the future.
\subsection{Scale-Coupling Factor for Linking the Astronomical and Cosmochemical Abundance Scales}

For a complete set of solar system elemental abundances, the solar and meteoritic abundances need to be combined and a scale-coupling factor is needed to combine the meteoritic (cosmochemical) and astronomical (solar) abundance scales. The linear cosmochemical scale is often used for the non-volatile elements retained in meteorites, and in this scale abundances are usually given relative to $10^6$ Si atoms. The logarithmic atomic astronomical scale, normalized to $\log_{10} = 12$ for H atoms, is used for all elements that can be quantitatively measured in the solar atmosphere, which includes the highly volatile elements that were lost from meteorites. 

One could just combine the solar and meteoritic values by normalizing each data set to a common, well-determined element such as Si or Mg and then combine the data into one set. However, then everything relies on a single element, and each time a revision in the abundance of the reference element the entire abundance scale has to be re-calibrated. Therefore it is more practical to use a set of elements to anchor the astronomic and meteoritic scales, which makes the scaling more robust.

\begin{figure}[t]
\centering
\includegraphics[width=0.8\textwidth]{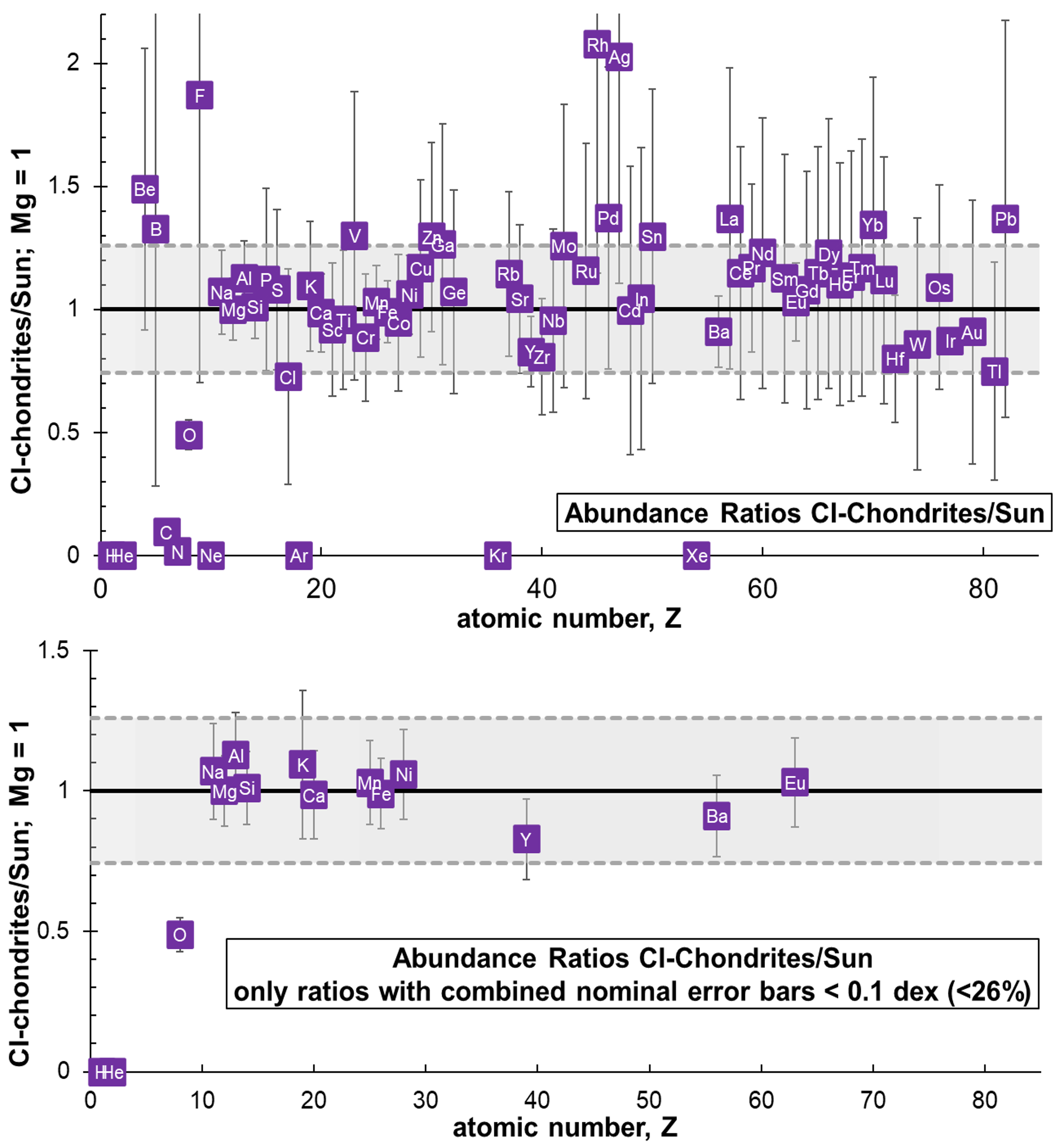}
\caption{Meteoritic to photospheric abundance ratios as a function of atomic number. The grey band spans $\pm 26 \%$ uncertainty from agreement at unity (solid line). Top: All elements for which the comparison can be made. Bottom: Only elements where the combined nominal error bars are below 0.1 dex (26\%). The elements that agree well include those that are well determined in the Sun by full 3D NLTE analyses.}
\label{chap1:fig6}
\end{figure}

We derive the scale-coupling-factor from the abundances of Na, Mg, Al, Si, Ca, Mn, Fe, and Ni in CI-chondrites and the Sun. These elements have the best 3D NLTE analyses in the photosphere (see Table \ref{tab:1} and Figure \ref{chap1:fig6}), are well determined in CI-chondrites, and are more abundant among the elements that are not highly volatile. For calculating the scale-coupling factor, the meteoritic mass concentration values for each element $i$ are normalized to Si $= 10^6$ atoms and converted to $\log_{10}$-values, thus giving
\begin{align}\label{chap1:eq6}
 \log_{10} N(i) (\rm{CI}) = 6 + \log_{10} [{(c(i) /atwt(i)} / {c(\rm{Si})/atwt(\rm{Si})}], 
\end{align}
where $c(i)$ is the concentration of the element $i$ in ppm by mass.

Then the difference to the solar abundances, $A(i)_{\rm photosphere} - \log_{10} N(i)_{\rm CI}$, gives the scaling factor (a constant in $\log$ space) for each element that is used for determining the factor. The scale-coupling factor from the average of the eight elements above is $1.551 \pm 0.020$. The factor can be applied to either solar or meteoritic abundances, depending on application. For bringing CI-chondrite values to the photospheric abundance scale we get
\begin{align}\label{chap1:eq7}
A(i)_{\rm CI} = \log_{10} N(i)_{\rm CI} + 1.551
\end{align}

For direct comparison of the photospheric abundances to the meteoritic abundance scale based on Si, the conversion is 

\begin{align}\label{chap1:eq8}
 N(i)_{\rm photosphere} = 10^{(A(i)_{\rm photosphere} – 1.551)}
\end{align}

These relationships were used to assemble the data collected in Table \ref{tab:1}. Figure \ref{chap1:fig6} shows the ratios of meteoritic to photospheric abundances as a function of atomic number. The top graph shows all elements, for which direct comparison can be made. Most elements agree within 26\%, which is the span of the grey band. Elements that deviate the most are the sunspot values for F, Cl, and Tl. Other deviant elements, such as Rh and Ag, are poorly known in the Sun. The disagreement of H, C, N, O and the noble gases is not surprising, because these elements are volatile or form volatile compounds that were not retained by meteorites. The bottom graph only shows elements where the combined nominal error bars are below 0.1 dex (26\%). The elements that agree well include those that are determined with superior 3D NLTE analyses in the solar photosphere.

\subsection{Gravitational Settling of Heavy Elements from the Photosphere and Protosolar Metallicity}

During the Sun’s lifetime, heavy elements gravitationally settled towards the solar interior and the convection zone. The photosphere became depleted in heavy elements relative to hydrogen. This process is somewhat counteracted by radiative levitation, but the latter cannot prevent it. 

Thus, the present-day photospheric abundances relative to H are \textbf{smaller} than at the time when the Sun formed. For obtaining the original, proto-solar ($=$ solar system) abundances, settling correction factors, SF, are used to correct for the 10-20\% (model-dependent) reductions in element/H ratios. The current uncertainties in elemental abundances do not indicate discernible effects on settling efficiency as a function of atomic mass; moreover, one would need to determine how such a baseline for gravitationally undisturbed or non-levitated abundances would be defined to begin with. The comparison of meteoritic and photospheric abundances in Figure \ref{chap1:fig6} might suggest some mass fractionation, because many of the heaviest elements are higher in the photosphere (or depleted in CI-chondrites). However, such a comparison is difficult because the analytical methods are not fully optimized yet and/or uncertainties are still too large for solar photospheric as well as for meteoritic abundances of many elements, which are involved in defining the apparent fractionations.

The settling corrections or factors (“SF”) can only assume that all elements heavier than lithium were reduced in the convective envelope by the same factor over time (but see also \citealt{Piersanti2007} for settling factors computed for each element). The settling factors used here are $\log_{10}$(SF(He)) $= 0.070$ dex ($17.5$\% change) and $\log_{10}$(SF(Li-U)) $= 0.0882$ dex ($22.5$\% change) for the heavier elements and are based on models by \citet{Yang2019}. However, we note that these SF values are based on a solar model with an ad-hoc scaling applied to enhance the efficiency of diffusion and gravitational settling. A more comprehensive value would be needed for the new Z/X values, which are used in this chapter and in \citet{Lodders2025}. For more details and discussion, see \citet{Lodders2021}. In order to obtain the proto-solar values, the $\log_{10}$ of the settling correction is applied to the present-day abundances:

\begin{align}\label{chap1:eq9}
  A(i)_{\rm proto-solar} = 12 + \log_{10}(N_i/N_{\rm H})_{\rm proto-solar} = A(i)_{\rm present} + \log_{10}{\rm SF} =\\
  12 + \log_{10}(N_i/N_{\rm{H}})_{\rm present}+ \log_{10}{\rm SF}
\end{align}

The protosolar mass fractions ($X_0,Y_0,Z_0$ indicated by the subscript “0”) are then obtained from the present-day solar ratios using the linear (not log scale) SF as:
\begin{align}\label{chap1:eq10}
 Y_0/X_0 = (Y/X) / \rm{SF}(\rm{He}),
\end{align}

\begin{align}\label{chap1:eq11}
 Z_0/X_0 = (Z/X) / \rm{SF}(\rm{Li-U}).
\end{align}

The atomic weights (necessary to obtain mass fractions from atomic elemental abundances, see above) cancel out when present-day and protosolar ratios are related. Using the results for the ratios and the mass-balance relation, $X_0 = 1/(1 + Y_0/X_0 + Z_0/X_0)$, the proto-solar mass fractions listed in Table \ref{tab:3} are derived. Note that the depletion factors listed in \citet{Piersanti2007} are not equal to the settling factors defined above; their factors correspond to the ratios of the absolute mass fraction ratios such as $(X_0-X)/X_0$, $(Y_0-Y)/Y_0$, $(Z_0-Z)/Z_0$.

The presence of long-lived (above the lifetime of the current solar system age) radioactive isotopes requires additional adjustments, when calculating protosolar abundances. The abundances of radioactive parent isotopes were adjusted for decay loss over time and the stable daughter isotopes for gain. 

The protosolar elemental abundances - representing the solar system abundances - are listed in Table \ref{tab:4} on the astronomical (logarithmic) abundance scale with $A(\rm{H}) = 12$, on the linear cosmochemical abundance scale with Si $= 10^6$ atoms, and on a concentration scale by mass (as mass fractions).

\subsection{Isotopic Composition and Nuclide Abundances}

\begin{figure}[t]
\centering
\includegraphics[width=0.8\textwidth]{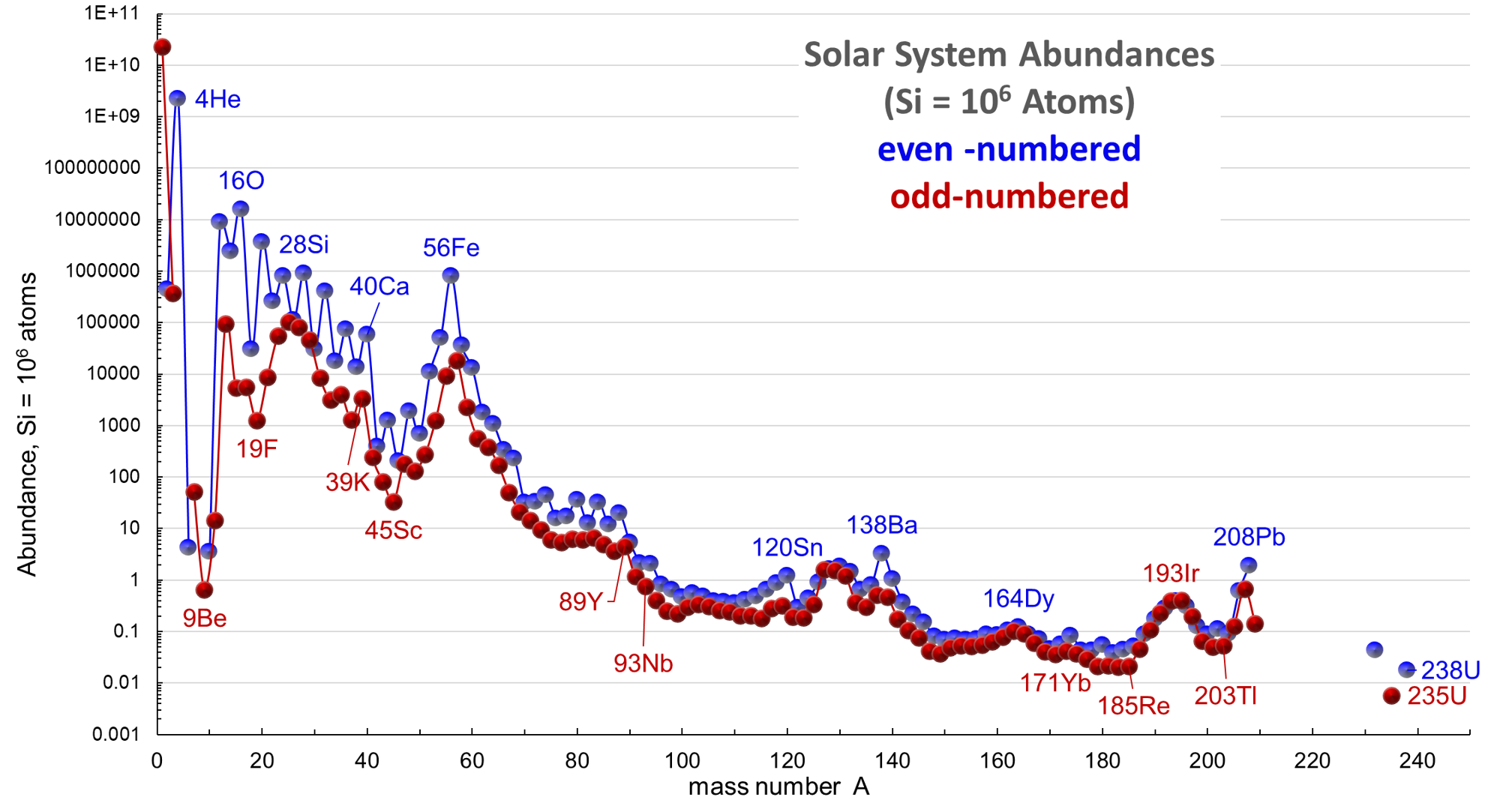}
\caption{The isotopic (or nuclide) abundances as a function of mass number. As in Figure 1 for the elements with atomic number, there is an odd-even effect on nuclide abundances with nuclide mass numbers. Nuclide abundances that fall along isobars are summed up here (isobars = nuclides with the same mass numbers (=sum of a nuclide’s protons and neutrons) but different atomic (proton) numbers because neutron and proton numbers vary).}
\label{chap1:fig7}
\end{figure}

We only briefly comment on the nuclide abundances listed in Table \ref{tab:5} and shown in Figure \ref{chap1:fig7}. The nuclide abundances are calculated from the elemental proto-solar abundances in Table \ref{tab:5} using the isotopic composition of the elements in atom-percent. The isotopic composition of the elements is from the review by \citet{Meija2016} except for H, the noble gases, C, N, and O, which are taken from the following studies. The proto-solar D/H ratio $(1.97(\pm  0.35) \times 10^{–5}$ is from \citet{Geiss2003}. The $^{3}$He/$^{4}$He of $1.66(\pm 0.05) \times 10^{–4}$ of Jupiter’s atmosphere is adopted as proto-solar value \citep{Mahaffy1998, Geiss2003}. We use $^{14}$N/$^{15}$N $= 442$ from the solar wind, and this value is similar to that found for Jupiter \citep{Abbas2004, Marty2010, Fueri2015}. The adopted oxygen isotopic composition of the solar wind is lighter (there is more of the light $^{16}$O than the isotopically heavier $^{17}$O and $^{18}$O) by about 7 percent than O in the Earth and in most meteoritic materials \citep{McKeegan2011, Laming2017b}. The C-isotopic composition is still relatively close to the terrestrial and meteoritic values. The noble gases (except He) are also based on the solar wind data from \citet{Pepin2012}, \citet{Meshik2014}, and \citet{Meshik2020}. 

Terrestrial isotopic abundance ratios are used, because information about the isotopic composition of elements - other than the highly volatile elements H, C, N, O, and the noble gases - in the photosphere and/or solar wind is very limited. The isotopic composition of several stable elements in different meteorite groups and planetary material has been studied extensively for many elements (e.g., review by \citealt{Teng2017}). One reason why the terrestrial isotopic compositions of rock-forming elements (other than oxygen) can be used is because meteoritic and planetary materials are typically in agreement to within per-mil levels or less with terrestrial values. For the purpose of comparing the absolute abundance distribution of the isotopes (or nuclides) in Figure \ref{chap1:fig7} (similar to what was done in Table \ref{tab:1} and Figure \ref{chap1:fig1} for the elements), it does not matter much whether terrestrial or meteoritic isotopic compositions are used. This is because the solar system materials are relatively homogeneous at a 1\% level (again excluding the highly volatile elements; e.g. \citealt{Birck2004}). The isotopic variations in meteorites and planetary materials are much smaller than the uncertainties associated with the elemental abundances, and the nuclide distribution in Figure \ref{chap1:fig7} is then much more sensitive to elemental variations than to the isotopic variations of a given element.

Ideally one would prefer to use solar or CI-chondrites for the isotopic composition of rocky elements. However, there is another unresolved issue. While isotopic compositional data exist for several elements relative to some laboratory or terrestrial rock isotopic standard(s), the absolute isotopic compositions of the elements cannot be computed because the absolute composition of the standard(s) is unknown (i.e., the standard is not calibrated against a known absolute composition such as given for the IUPAC values in \citealt{Meija2016}). 

Another issue concerns the highly volatile elements, for which mainly solar wind isotopic compositions are used. The solar wind appears isotopically light (i.e., the lighter isotopes of N, O, the noble gases, and probably other elements are more abundant) compared to the Earth and other planetary rocky materials and meteorites. Isotope fractionations can also occur during solar wind formation, compounding the problem. But the observed differences in e.g., O (7\% from solar wind to terrestrial) are intrinsic and are unlikely to only stem from isotope fractionation during formation of the solar wind. If the solar wind is representative of the isotopic composition of the Sun, and the solar system as a whole, the reasons for the differences between the isotopically light Sun and the isotopically heavy planets, moons, asteroids, and comets are puzzling. These differences await more explanations. 

The abundance distribution of the isotopes as a function of mass number in Figure \ref{chap1:fig7} reveals the high abundance peak in the iron region (“Fe-peak”). This is due to the increasing binding energy ($=$ higher stability) per nucleus up to iron and nickel nuclides. Starting with H, energy is released during formation of nuclei by nuclear fusion of the elements up to the Fe-peak. Elements beyond the Fe-peak are made by neutron capture nuclear reactions, which require energy input. The two principal neutron capture processes also leave their signatures as “peaks” in the nuclide distribution. Elements built by the slow-neutron capture process (slow compared to beta-decay lifetimes of the intermediate nuclides involved) are abundant in three mass number regions: around A $= 90–100$ (Sr, Y, Zr), around A $= 130–140$ (Ba, La, Ce), and around A $=208$ in the Pb region. The rapid neutron capture process (r-process) leaves its major mark in a heightened and broad peak around 180 – 200 (Os, Ir, Pt) but also around A $= 130$ (Te, Xe) just before the second s-process peak. The peaks are in regions where nuclides have so-called “magic numbers” of neutrons (i.e., 50, 82, or 126), that is, closed nuclear shells. Such nuclides have small neutron-capture cross-sections, and therefore are more stable. Hence, their abundances are larger at these mass numbers. The elemental (Figure \ref{chap1:fig1}) and nuclide abundance distributions (Figure \ref{chap1:fig7}) that stimulated ideas about nuclear structure and nucleosynthesis in stars since about a century ago remain powerful tests for our understanding about the origin of the elements. 
\section{Outlook on Solar Abundance Measurements}

Whereby enormous progress has been achieved in the solar photospheric abundance diagnostic over the past decade, many unresolved issues still remain. This includes observational and theoretical limitations that need to be overcome to reach a percent precision quality of the solar composition. On the one hand, the current data are often limited by the complexity of observed data: continuum normalization and line blending are among the most challenging aspects, which for many elements especially those with lines in the blue and near-UV range, constitute a significant source of uncertainty. Whereas strict pre-selection and filtering of lines has been a common practice so far, future work will have to consider a more complete analysis of the solar spectrum, including weak and blended features, but also stronger features with more accurate atomic and molecular data, in order to avoid biasing the results by subjective choices. In parallel, future developments in solar abundance determinations will progressively rely on new high-resolution spatially-resolved solar spectra \citep{Reiners2016}, in order to test and validate the abundance results obtained at differentiating pointings across the disc of the Sun.

Improvements of models combining 3D atmospheric solar models and non-local-thermodynamic equilibrium (NLTE) radiative transfer will lead to better descriptions of the solar structure and composition. The availability of different 3D RHD simulations relying on different micro-physics will help to understand the differences between the current results. A key physical concept that is currently missing in most solar measurements is the solar chromosphere and its effect on the photospheric structure, and hence on elemental abundances. The first limited results from MHD-corrected solar abundance studies are available \citep{Bergemann2021}, but such work has to be done for all elements from first principles, to ensure robust results. Atomic and molecular data provide the basis, and zero-point for accurate abundance calculations and more progress on key quantities, especially the cross-sections for photo-excitation and ionizations \citep[e.g.,][]{Nahar2015, Bautista2022}, but also critically photo-dissociation energy-resolved data \citep{Hrodmarsson2023}, along with data for collisionally induced reactions \citep{Belyaev2012, Barklem2016} is essential for further quantitative solar work. NLTE modelling is currently becoming the main-stream, but for many trans-Fe elements such work remains to be done, with the primary focus being on developing and validating 3D NLTE calculations for molecular lines in the solar spectrum, including the key C- and O-bearing species. Molecular lines provide important complementary estimates of elemental abundances, but they are also the unique tracer of detailed isotopic ratios, allowing us to independently test and validate isotopic constraints from other solar system bodies. Finally, exciting progress will come from more integrated models and analyses \citep{Truong2024} that utilize constraints from helioseismology, spectroscopy, meteoritics, and planetary science, including results from space missions. 
\section{Conclusion/Summary}

The photospheric chemical composition of the Sun represents a fundamental quantity in astronomy and astrophysics. It is used as input in a diversity of models, including stellar atmospheres, solar and stellar structures and evolution, planet structure, growth of protoplanetary disks, interstellar matter, and chemical evolution of stellar populations and galaxies. It is also used as zero-point for all abundance measurements in astronomy.

Major progress in understanding the solar composition has been achieved over the past decade thanks to impressive advances in theoretical calculations of the solar atmospheric structure (by means of 3D RHD), detailed radiation transfer in the atmosphere (by means of NLTE), and improvements in atomic and molecular data (by means of laboratory experiment and quantum mechanics calculations). Advances in understanding the physics of solar spectrum have been further accompanied by improvements in the quality and dimensionality of solar observations (spectra taken at different positions across the solar disc), which are increasingly used for validation of measurements. \textbf{The solar atmospheric abundances represent, arguably, the most robust and diverse set of chemical composition measurements in astronomy.} In future, integration of magneto-hydrodynamics, especially for magnetically active regions, NLTE analyses for molecules, and tighter constraints on isotopic abundance ratios, will represent a qualitative step forward.

Solar photospheric composition represents a major observational constraint on models of the interior structure and evolution of the Sun. Recent 3D NLTE measurements by two independent groups arrive at different conclusions, regarding the key chemical elements and bulk solar metallicity. The low-metallicity solar abundances \citep{Asplund2021} present a problem for the models of solar interior \citep{Villante2014}. In contrast, the high-metallicity solar abundances \citep{Bergemann2021, Magg2022} lead to largely consistent predictions of the Standard Solar Models and helioseismology \citep{Basinger2024, Yang2024}, are consistent with measurements based on solar neutrino fluxes \citep{Appel2022, Basilico2023}, and with combined analyses of solar wind and solar system data \citep{Truong2024}.
%

\input{latex_table_1}
\input{latex_table_2}
\input{latex_table_3}

\input{latex_table_4}
\input{latex_table_5}

\bibliographystyle{Harvard}
\bibliography{reference}

\end{document}

%% file: latex_table_1.tex
%
%
%
%
\begin{ThreePartTable}
    \renewcommand{\LTcapwidth}{\linewidth}
    \begin{longtable}{ccccccccccccc}
        \caption{\label{tab:1}Solar Photospheric Abundances and Meteoritic Abundances from CI -chondrites.}\\
        \toprule
        Z & E & \multicolumn{2}{c}{Sun Convection Zone} & \multicolumn{2}{c}{CI-Chondrites} & \multicolumn{3}{c}{Sun Convection Zone} & \multicolumn{3}{c}{CI-Chondrites} & CI/Solar \\
         &  & \multicolumn{2}{c}{(mainly photosphere)} &  &  & \multicolumn{3}{c}{(mainly photosphere)} &  &  &  &  \\
         &  & A(E) & $\pm\sigma$ & A(E)$^{*}$ & $\pm\sigma$ & N(E)$^{**}$ & $\pm\sigma$ & $\pm\sigma$ & N(E)$^{***}$ & $\pm\sigma$ & $\pm\sigma$ &  \\
         &  & dex & dex & dex & dex &  &  & \% &  &  & \% &  \\
        \midrule
        \endfirsthead
        \caption{continued.}\\
        \toprule
        Z & E & \multicolumn{2}{c}{Sun Convection Zone} & \multicolumn{2}{c}{CI-Chondrites} & \multicolumn{3}{c}{Sun Convection Zone} & \multicolumn{3}{c}{CI-Chondrites} & CI/Solar \\
         &  & \multicolumn{2}{c}{(mainly photosphere)} &  &  & \multicolumn{3}{c}{(mainly photosphere)} &  &  &  &  \\
         &  & A(E) & $\pm\sigma$ & A(E) & $\pm\sigma$ & N(E)$^{**}$ & $\pm\sigma$ & $\pm\sigma$ & N(E)$^{***}$ & $\pm\sigma$ & $\pm\sigma$ &  \\
         &  & dex & dex & dex$^{*}$ & dex &  &  & \% &  &  & \% &  \\
        \midrule
        \endhead
        \bottomrule
        \endfoot
        1 & H & 12 & 0.004 & 8.24 & 0.04 & 2.81E+10 & 3.00E+08 & 0.9 & 4.86E+06 & 4.49E+05 & 9.2 & 1.70E-04 \\
        2 & He & 10.922 & 0.012 & 1.33 & 0.04 & 2.35E+09 & 6.60E+07 & 2.8 & 0.604 & 0.06 & 10 & 2.60E-10 \\
        3 & Li & 1.04 & 0.09 & 3.3 & 0.02 & 0.308 & 0.071 & 23 & 56.2 & 2.7 & 4.7 & 182 \\
        4 & Be & 1.21 & 0.14 & 1.37 & 0.02 & 0.456 & 0.174 & 38 & 0.658 & 0.038 & 5.8 & 1.4 \\
        5 & B & 2.7 & 0.25 & 2.81 & 0.05 & 14.1 & 11 & 77.8 & 18.1 & 2.3 & 12.8 & 1.3 \\
        6 & C & 8.51 & 0.09 & 7.47 & 0.07 & 9.10E+06 & 2.10E+06 & 23 & 8.29E+05 & 1.45E+05 & 17.5 & 0.1 \\
        7 & N & 7.94 & 0.11 & 6.12 & 0.17 & 2.45E+06 & 7.10E+05 & 28.8 & 36956 & 18153 & 49.1 & 1.50E-02 \\
        8 & O & 8.76 & 0.05 & 8.44 & 0.01 & 1.62E+07 & 2.00E+06 & 12.2 & 7.67E+06 & 1.32E+05 & 1.7 & 0.5 \\
        9 & F & 4.4 & 0.2 & 4.66 & 0.09 & 706 & 413 & 58.5 & 1280 & 277 & 21.7 & 1.8 \\
        10 & Ne & 8.15 & 0.12 & -1.08 & 0.04 & 3.97E+06 & 1.26E+06 & 31.8 & 2.36E-03 & 2.36E-04 & 10 & 5.90E-10 \\
        11 & Na & 6.29 & 0.05 & 6.31 & 0.04 & 54838 & 6691 & 12.2 & 56838 & 5868 & 10.3 & 1 \\
        12 & Mg & 7.58 & 0.05 & 7.57 & 0.01 & 1.07E+06 & 1.30E+05 & 12.2 & 1.04E+06 & 3.25E+04 & 3.1 & 1 \\
        13 & Al & 6.43 & 0.05 & 6.47 & 0.02 & 75698 & 9237 & 12.2 & 82707 & 4687 & 5.7 & 1.1 \\
        14 & Si & 7.56 & 0.05 & 7.55 & 0.02 & 1.02E+06 & 1.20E+05 & 12.2 & 1.00E+06 & 4.13E+04 & 4.1 & 1 \\
        15 & P & 5.44 & 0.12 & 5.48 & 0.04 & 7746 & 2465 & 31.8 & 8413 & 757 & 9 & 1.1 \\
        16 & S & 7.16 & 0.11 & 7.18 & 0.04 & 406521 & 117180 & 28.8 & 425635 & 37798 & 8.9 & 1 \\
        17 & Cl & 5.43 & 0.2 & 5.28 & 0.06 & 7570 & 4428 & 58.5 & 5326 & 817 & 15.3 & 0.7 \\
        18 & Ar & 6.5 & 0.12 & -0.46 & 0.04 & 88937 & 28305 & 31.8 & 9.66E-03 & 9.66E-04 & 10 & 1.10E-07 \\
        19 & K & 5.09 & 0.09 & 5.12 & 0.03 & 3460 & 797 & 23 & 3667 & 276 & 7.5 & 1.1 \\
        20 & Ca & 6.35 & 0.06 & 6.33 & 0.03 & 62963 & 9328 & 14.8 & 60135 & 3641 & 6.1 & 1 \\
        21 & Sc & 3.13 & 0.11 & 3.08 & 0.03 & 37.9 & 10.9 & 28.8 & 33.8 & 2.2 & 6.4 & 0.9 \\
        22 & Ti & 4.97 & 0.11 & 4.94 & 0.03 & 2625 & 757 & 28.8 & 2433 & 165 & 6.8 & 0.9 \\
        23 & V & 3.89 & 0.16 & 3.99 & 0.03 & 218 & 97 & 44.5 & 275 & 20 & 7.3 & 1.3 \\
        24 & Cr & 5.74 & 0.11 & 5.67 & 0.02 & 15456 & 4455 & 28.8 & 13255 & 588 & 4.4 & 0.9 \\
        25 & Mn & 5.52 & 0.05 & 5.52 & 0.03 & 9313 & 1136 & 12.2 & 9282 & 734 & 7.9 & 1 \\
        26 & Fe & 7.51 & 0.05 & 7.49 & 0.01 & 910088 & 111048 & 12.2 & 872789 & 30194 & 3.5 & 1 \\
        27 & Co & 4.95 & 0.11 & 4.91 & 0.02 & 2507 & 723 & 28.8 & 2297 & 116 & 5.1 & 0.9 \\
        28 & Ni & 6.24 & 0.06 & 6.25 & 0.01 & 48875 & 7241 & 14.8 & 50184 & 1481 & 3 & 1 \\
        29 & Cu & 4.24 & 0.11 & 4.29 & 0.05 & 489 & 141 & 28.8 & 552 & 62 & 11.3 & 1.1 \\
        30 & Zn & 4.55 & 0.11 & 4.65 & 0.03 & 998 & 288 & 28.8 & 1251 & 89 & 7.1 & 1.3 \\
        31 & Ga & 3.02 & 0.14 & 3.11 & 0.03 & 29.4 & 11.2 & 38 & 36.1 & 2.6 & 7.1 & 1.2 \\
        32 & Ge & 3.62 & 0.14 & 3.64 & 0.03 & 117 & 45 & 38 & 122 & 8 & 6.9 & 1 \\
        33 & As & \ldots & \ldots & 2.34 & 0.04 & \ldots & \ldots & \ldots & 6.16 & 0.6 & 9.7 & \ldots \\
        34 & Se & \ldots & \ldots & 3.41 & 0.01 & \ldots & \ldots & \ldots & 71.7 & 2.3 & 3.3 & \ldots \\
        35 & Br & \ldots & \ldots & 2.65 & 0.09 & \ldots & \ldots & \ldots & 12.4 & 3 & 23.9 & \ldots \\
        36 & Kr & 3.31 & 0.12 & -2.23 & 0.04 & 57.4 & 18.3 & 31.8 & 1.64E-04 & 1.64E-05 & 10 & 2.90E-06 \\
        37 & Rb & 2.35 & 0.11 & 2.39 & 0.02 & 6.3 & 1.81 & 28.8 & 6.98 & 0.37 & 5.3 & 1.1 \\
        38 & Sr & 2.93 & 0.11 & 2.93 & 0.01 & 23.9 & 6.9 & 28.8 & 24.2 & 0.8 & 3.4 & 1 \\
        39 & Y & 2.3 & 0.06 & 2.2 & 0.04 & 5.61 & 0.83 & 14.8 & 4.5 & 0.4 & 8.8 & 0.8 \\
        40 & Zr & 2.68 & 0.11 & 2.57 & 0.02 & 13.5 & 3.9 & 28.8 & 10.5 & 0.6 & 5.8 & 0.8 \\
        41 & Nb & 1.47 & 0.14 & 1.44 & 0.03 & 0.83 & 0.316 & 38 & 0.767 & 0.063 & 8.2 & 0.9 \\
        42 & Mo & 1.88 & 0.16 & 1.97 & 0.04 & 2.13 & 0.95 & 44.5 & 2.6 & 0.27 & 10.6 & 1.2 \\
        43 & Tc & \ldots & \ldots & \ldots & \ldots & \ldots & \ldots & \ldots & \ldots & \ldots & \ldots & \ldots \\
        44 & Ru & 1.75 & 0.16 & 1.8 & 0.03 & 1.58 & 0.7 & 44.5 & 1.77 & 0.11 & 6.2 & 1.1 \\
        45 & Rh & 0.78 & 0.16 & 1.08 & 0.02 & 0.169 & 0.075 & 44.5 & 0.341 & 0.015 & 4.5 & 2 \\
        46 & Pd & 1.57 & 0.16 & 1.69 & 0.02 & 1.04 & 0.47 & 44.5 & 1.39 & 0.06 & 4.6 & 1.3 \\
        47 & Ag & 0.96 & 0.16 & 1.25 & 0.04 & 0.256 & 0.114 & 44.5 & 0.504 & 0.046 & 9.2 & 2 \\
        48 & Cd & 1.77 & 0.2 & 1.75 & 0.03 & 1.66 & 0.97 & 58.5 & 1.6 & 0.11 & 6.9 & 1 \\
        49 & In & 0.8 & 0.2 & 0.8 & 0.03 & 0.177 & 0.104 & 58.5 & 0.179 & 0.013 & 7 & 1 \\
        50 & Sn & 2.02 & 0.16 & 2.12 & 0.05 & 2.94 & 1.31 & 44.5 & 3.7 & 0.44 & 12 & 1.3 \\
        51 & Sb & \ldots & \ldots & 1.08 & 0.07 & \ldots & \ldots & \ldots & 0.34 & 0.056 & 16.6 & \ldots \\
        52 & Te & \ldots & \ldots & 2.23 & 0.02 & \ldots & \ldots & \ldots & 4.75 & 0.19 & 3.9 & \ldots \\
        53 & I & \ldots & \ldots & 1.76 & 0.15 & \ldots & \ldots & \ldots & 1.6 & 0.64 & 40.2 & \ldots \\
        54 & Xe & 2.3 & 0.12 & -1.91 & 0.04 & 5.61 & 1.79 & 31.8 & 3.49E-04 & 3.50E-05 & 10 & 6.20E-05 \\
        55 & Cs & \ldots & \ldots & 1.12 & 0.03 & \ldots & \ldots & \ldots & 0.368 & 0.024 & 6.5 & \ldots \\
        56 & Ba & 2.27 & 0.06 & 2.22 & 0.02 & 5.24 & 0.78 & 14.8 & 4.62 & 0.27 & 5.8 & 0.9 \\
        57 & La & 1.1 & 0.16 & 1.22 & 0.02 & 0.354 & 0.158 & 44.5 & 0.47 & 0.019 & 4 & 1.3 \\
        58 & Ce & 1.58 & 0.16 & 1.63 & 0.02 & 1.07 & 0.48 & 44.5 & 1.19 & 0.05 & 4.2 & 1.1 \\
        59 & Pr & 0.75 & 0.11 & 0.8 & 0.02 & 0.158 & 0.046 & 28.8 & 0.179 & 0.008 & 4.4 & 1.1 \\
        60 & Nd & 1.42 & 0.16 & 1.5 & 0.02 & 0.74 & 0.33 & 44.5 & 0.88 & 0.046 & 5.2 & 1.2 \\
        61 & Pm & \ldots & \ldots & \ldots & \ldots & \ldots & \ldots & \ldots & \ldots & \ldots & \ldots & \ldots \\
        62 & Sm & 0.95 & 0.16 & 0.99 & 0.02 & 0.251 & 0.112 & 44.5 & 0.273 & 0.014 & 5.2 & 1.1 \\
        63 & Eu & 0.57 & 0.06 & 0.57 & 0.02 & 0.104 & 0.015 & 14.8 & 0.104 & 0.005 & 4.5 & 1 \\
        64 & Gd & 1.08 & 0.16 & 1.1 & 0.02 & 0.338 & 0.151 & 44.5 & 0.353 & 0.017 & 4.9 & 1 \\
        65 & Tb & 0.31 & 0.16 & 0.36 & 0.02 & 0.0574 & 0.0256 & 44.5 & 0.0638 & 0.0028 & 4.3 & 1.1 \\
        66 & Dy & 1.1 & 0.16 & 1.17 & 0.02 & 0.354 & 0.158 & 44.5 & 0.421 & 0.018 & 4.2 & 1.2 \\
        67 & Ho & 0.48 & 0.16 & 0.51 & 0.02 & 0.0849 & 0.0378 & 44.5 & 0.0908 & 0.0037 & 4 & 1.1 \\
        68 & Er & 0.93 & 0.16 & 0.97 & 0.02 & 0.239 & 0.107 & 44.5 & 0.263 & 0.011 & 4.3 & 1.1 \\
        69 & Tm & 0.11 & 0.16 & 0.16 & 0.02 & 0.0362 & 0.0161 & 44.5 & 0.041 & 0.0019 & 4.6 & 1.1 \\
        70 & Yb & 0.85 & 0.16 & 0.96 & 0.02 & 0.199 & 0.089 & 44.5 & 0.259 & 0.01 & 4 & 1.3 \\
        71 & Lu & 0.1 & 0.16 & 0.13 & 0.02 & 0.0354 & 0.0158 & 44.5 & 0.0384 & 0.0022 & 5.7 & 1.1 \\
        72 & Hf & 0.86 & 0.12 & 0.75 & 0.02 & 0.204 & 0.065 & 31.8 & 0.158 & 0.009 & 5.8 & 0.8 \\
        73 & Ta & \ldots & \ldots & -0.11 & 0.02 & \ldots & \ldots & \ldots & 0.0218 & 0.001 & 4.7 & \ldots \\
        74 & W & 0.79 & 0.2 & 0.71 & 0.05 & 0.173 & 0.101 & 58.5 & 0.144 & 0.017 & 11.9 & 0.8 \\
        75 & Re & \ldots & \ldots & 0.3 & 0.02 & \ldots & \ldots & \ldots & 0.0564 & 0.0023 & 4.1 & \ldots \\
        76 & Os & 1.36 & 0.14 & 1.38 & 0.01 & 0.644 & 0.245 & 38 & 0.68 & 0.019 & 2.8 & 1.1 \\
        77 & Ir & 1.42 & 0.2 & 1.35 & 0.02 & 0.74 & 0.433 & 58.5 & 0.625 & 0.037 & 5.9 & 0.8 \\
        78 & Pt & \ldots & \ldots & 1.64 & 0.03 & \ldots & \ldots & \ldots & 1.22 & 0.09 & 7.8 & \ldots \\
        79 & Au & 0.91 & 0.2 & 0.85 & 0.04 & 0.229 & 0.134 & 58.5 & 0.201 & 0.018 & 8.8 & 0.9 \\
        80 & Hg & \ldots & \ldots & 1.14 & 0.13 & \ldots & \ldots & \ldots & 0.385 & 0.139 & 36.1 & \ldots \\
        81 & Tl & 0.95 & 0.2 & 0.81 & 0.04 & 0.251 & 0.147 & 58.5 & 0.182 & 0.018 & 9.9 & 0.7 \\
        82 & Pb & 1.95 & 0.2 & 2.07 & 0.03 & 2.51 & 1.47 & 58.5 & 3.32 & 0.241 & 7.3 & 1.3 \\
        83 & Bi & \ldots & \ldots & 0.7 & 0.03 & \ldots & \ldots & \ldots & 0.142 & 0.009 & 6.2 & \ldots \\
        90 & Th & \ldots & \ldots & 0.09 & 0.04 & \ldots & \ldots & \ldots & 0.0345 & 0.0037 & 10.9 & \ldots \\
        92 & U & \ldots & \ldots & -0.5 & 0.04 & \ldots & \ldots & \ldots & 0.00896 & 0.00085 & 9.5 & \ldots \\      
    \end{longtable}
    \begin{tablenotes}
        \small
        \item $^{*}$CI-chondrites: A(E) = 1.551+log N(E) dex. $^{**}$ Solar: N(E) = $10^{A(X) - 1.551}$. $^{***}$ CI-chondrites: N(Si) = $10^6$ atoms.
    \end{tablenotes}
\end{ThreePartTable}

%% file: latex_table_2.tex
\renewcommand{\LTcapwidth}{\linewidth}
\begin{longtable}{ccc ccc}
    \caption{\label{tab:2} Elemental Abundances in Solar Corpuscular Radiation}\\
    \toprule
     Z & E & \multicolumn{2}{c}{Solar Energetic Particles} & \multicolumn{2}{c}{Bulk Solar Wind (Genesis)}  \\
     &     & \multicolumn{2}{c}{ Reames (2018) }           & \multicolumn{2}{c}{ Heber et al. (2021)}  \\
     &     & Mg $=100$ & $\pm\sigma$ &  Mg $=100$ & $\pm\sigma$  \\
    \midrule
    \endhead
    \bottomrule
    \endfoot
     1 & H  & 8.99E+05 & 1.12E+05 & 9.45E+05 & 1.00E+05 \\
     2 & He & 5.11E+04 & 2.80E+03 & 4.79E+04 & 1.90E+03 \\
     6 & C  & 236 & 6 & 371 & 25 \\
     7 & N  & 72  & 4 &  71 & 8 \\
     8 & O  & 562 & 6 & 676 & 73 \\
     10 & Ne & 88.2 & 5.6 & 78.6 & 3.3 \\ 
     11 & Na & 5.8 & 0.6 & 4.6 & 0.6 \\
     12 & Mg & 100 & 2 & 100 & 5 \\ 
     13 & Al & 8.82 & 0.9 & 8.21 & 0.91 \\
     14 & Si & 84.8 & 2.2 & ... & ... \\ 
     15 & P & 0.37 & 0.1 & ... & ... \\ 
     16 & S & 14 & 1.1 & ... & ... \\ 
     17 & Cl & 0.13 & 0.06 & ... & ... \\
     18 & Ar & 2.42 & 0.22 & 2.05 & 0.08 \\
     19 & K & 0.31 & 0.08 & 0.25 & 0.06 \\ 
     20 & Ca & 6.18 & 0.56 & 6.8 & 0.4 \\ 
     22 & Ti & 0.19 & 0.06 & ... & ... \\ 
     24 & Cr & 1.18 & 0.17 & 1.34 & 0.05 \\
     26 & Fe & 73.6 & 3.4 & 79.2 & 4.4 \\ 
     28 & Ni & 3.6 & 0.34 & ... & ... \\ 
     30 & Zn & 0.06 & 0.01 & ... & ... \\
     36 & Kr & ... & ... & 1.27E-03 & 1.00E-04 \\
     54 & Xe & ... & ... & 2.54E-04 & 2.40E-05 \\
\end{longtable}
\begin{tablenotes}
\small
\item $^{*}$ Abundances of the elements in Solar Energetic Particles (SEP) from Reames (2018), and in the Solar Wind as determined by the GENESIS mission from Heber et al. (2021), see also Huss et al. (2020). Here Na, K, Fe are from Jurewicz et al. (2024).
\end{tablenotes}

%% file: latex_table_3.tex
\renewcommand{\LTcapwidth}{\linewidth}
\begin{longtable}{lll}
\caption{\label{tab:3} Mass Fractions for Solar System Composition*}\\
\toprule
Mass Fraction & Present-Day & Protosolar \\
\midrule
$X$   & 0.7389 $\pm$ 0.0068 ($\pm$0.9\%) & 0.7061 $\pm$ 0.0065 ($\pm$0.9\%) \\
$Y$   & 0.2451 $\pm$ 0.0069 ($\pm$2.8\%) & 0.2752 $\pm$ 0.0077 ($\pm$2.8\%) \\
$Z$   & 0.0160 $\pm$ 0.0013 ($\pm$8\%)   & 0.0187 $\pm$ 0.0015 ($\pm$8\%)   \\
$Z/X$ & 0.0216 $\pm$ 0.0017 ($\pm$8\%)   & 0.0265 $\pm$ 0.0021 ($\pm$8\%)   \\
\endhead
\bottomrule
\endfoot
\end{longtable}
\begin{tablenotes}
\small
\item $^{*}$ Composition derived from photospheric and CI-chondritic abundances. Mass fraction $X$ is for H, $Y$ for He, and $Z$ is for the sum of Li to U. See text.
\end{tablenotes}

%% file: latex_table_4.tex
\begin{ThreePartTable}
\renewcommand{\LTcapwidth}{\linewidth}
\begin{longtable}{ccccc ccccc}
\caption{\label{tab:4}Proto-Solar or Solar System Elemental Abundances}\\
        \toprule
        Z & E & $\log_{10} (\rm{X/H})+12$ & $\pm$ $\sigma$ & Si $= 10^{6}$ &  $\pm$ $\sigma$ & Mass fractions &  $\pm$ $\sigma$ & $\sigma$ \% & Note* \\
        \midrule
        \endfirsthead
        \caption{continued.}\\
        \toprule
        Z & E & $\log_{10} (\rm{X/H})+12$ & $\pm$ $\sigma$ & Si $= 10^{6}$ &  $\pm$ $\sigma$ & Mass fractions &  $\pm$ $\sigma$ & $\sigma$ \% & Note* \\
        \midrule
        \endhead
        \bottomrule
        \endfoot
%
1 & H & 12 & 0.004 & 2.29E+10 & 2.10E+08 & 0.7061 & 0.0065 & 0.92 & s \\
2 & He & 10.992 & 0.012 & 2.25E+09 & 6.00E+07 & 0.2752 & 0.0077 & 2.81 &  s, t  \\
3 & Li & 3.39 & 0.02 & 56.1 & 2.7 & 1.19E-08 & 6.00E-10 & 4.74 & m \\
4 & Be & 1.46 & 0.02 & 0.657 & 0.038 & 1.81E-10 & 1.00E-11 & 5.79 & m \\
5 & B & 2.9 & 0.05 & 18.1 & 2.3 & 5.98E-09 & 7.70E-10 & 12.79 & m \\
6 & C & 8.6 & 0.09 & 9.08E+06 & 2.10E+06 & 3.34E-03 & 7.70E-04 & 23.08 & s \\
7 & N & 8.03 & 0.11 & 2.44E+06 & 7.10E+05 & 1.05E-03 & 3.00E-04 & 28.89 & s \\
8 & O & 8.85 & 0.05 & 1.62E+07 & 2.00E+06 & 7.90E-03 & 9.70E-04 & 12.23 & s \\
9 & F & 4.75 & 0.09 & 1278 & 277 & 7.42E-07 & 1.61E-07 & 21.71 & m \\
10 & Ne & 8.24 & 0.12 & 3.96E+06 & 1.26E+06 & 2.44E-03 & 7.80E-04 & 31.89 &  s, t  \\
11 & Na & 6.39 & 0.01 & 55852 & 1414 & 3.93E-05 & 1.00E-06 & 2.53 & a \\
12 & Mg & 7.66 & 0.01 & 1.04E+06 & 2.00E+04 & 7.70E-04 & 1.70E-05 & 2.25 & a \\
13 & Al & 6.55 & 0.03 & 81102 & 4956 & 6.69E-05 & 4.10E-06 & 6.11 & a \\
14 & Si & 7.64 & 0.01 & 1.00E+06 & 1.00E+04 & 8.59E-04 & 1.30E-05 & 1.49 & a \\
15 & P & 5.56 & 0.04 & 8395 & 757 & 7.95E-06 & 7.20E-07 & 9.02 & m \\
16 & S & 7.27 & 0.04 & 424747 & 37798 & 4.17E-04 & 3.70E-05 & 8.9 & m \\
17 & Cl & 5.37 & 0.06 & 5315 & 817 & 5.76E-06 & 8.90E-07 & 15.38 & m \\
18 & Ar & 6.59 & 0.12 & 88751 & 28305 & 9.85E-05 & 3.10E-05 & 31.89 &  s, t  \\
19 & K & 5.2 & 0.03 & 3664 & 277 & 4.38E-06 & 3.30E-07 & 7.55 & m \\
20 & Ca & 6.42 & 0.01 & 60378 & 2000 & 7.40E-05 & 2.50E-06 & 3.31 & a \\
21 & Sc & 3.17 & 0.03 & 33.7 & 2.2 & 4.63E-08 & 3.00E-09 & 6.43 & m \\
22 & Ti & 5.03 & 0.03 & 2428 & 165 & 3.55E-06 & 2.40E-07 & 6.8 & m \\
23 & V & 4.08 & 0.03 & 274 & 20 & 4.27E-07 & 3.10E-08 & 7.35 & m \\
24 & Cr & 5.76 & 0.02 & 13227 & 588 & 2.10E-05 & 9.00E-07 & 4.44 & m \\
25 & Mn & 5.61 & 0 & 9272 & 22 & 1.56E-05 & 4.00E-08 & 0.23 & a \\
26 & Fe & 7.58 & 0.01 & 8.74E+05 & 2.60E+04 & 1.49E-03 & 5.00E-05 & 3.02 & a \\
27 & Co & 5 & 0.02 & 2292 & 116 & 4.13E-06 & 2.10E-07 & 5.07 & m \\
28 & Ni & 6.34 & 0.01 & 50026 & 926 & 8.98E-05 & 1.70E-06 & 1.85 & a \\
29 & Cu & 4.38 & 0.05 & 551 & 62 & 1.07E-06 & 1.20E-07 & 11.28 & m \\
30 & Zn & 4.74 & 0.03 & 1248 & 89 & 2.50E-06 & 1.80E-07 & 7.1 & m \\
31 & Ga & 3.2 & 0.03 & 36 & 2.6 & 7.67E-08 & 5.50E-09 & 7.14 & m \\
32 & Ge & 3.72 & 0.03 & 121 & 8 & 2.70E-07 & 1.90E-08 & 6.87 & m \\
33 & As & 2.43 & 0.04 & 6.15 & 0.6 & 1.41E-08 & 1.40E-09 & 9.72 & m \\
34 & Se & 3.49 & 0.01 & 71.5 & 2.3 & 1.73E-07 & 6.00E-09 & 3.26 & m \\
35 & Br & 2.73 & 0.09 & 12.4 & 3 & 3.03E-08 & 7.30E-09 & 23.91 & m \\
36 & Kr & 3.4 & 0.12 & 57.3 & 18.3 & 1.47E-07 & 4.70E-08 & 31.89 & t \\
37 & Rb & 2.49 & 0.02 & 7.09 & 0.38 & 1.85E-08 & 1.00E-09 & 5.31 & m \\
38 & Sr & 3.02 & 0.01 & 24 & 0.8 & 6.43E-08 & 2.20E-09 & 3.36 & m \\
39 & Y & 2.29 & 0.04 & 4.49 & 0.4 & 1.22E-08 & 1.10E-09 & 8.86 & m \\
40 & Zr & 2.66 & 0.02 & 10.5 & 0.6 & 2.93E-08 & 1.70E-09 & 5.82 & m \\
41 & Nb & 1.52 & 0.03 & 0.766 & 0.063 & 2.18E-09 & 1.80E-10 & 8.22 & m \\
42 & Mo & 2.05 & 0.04 & 2.6 & 0.27 & 7.62E-09 & 8.10E-10 & 10.58 & m \\
43 & Tc & ... & ... & ... & ... & ... & ... & ... & ... \\
44 & Ru & 1.89 & 0.03 & 1.77 & 0.11 & 5.46E-09 & 3.40E-10 & 6.24 & m \\
45 & Rh & 1.17 & 0.02 & 0.341 & 0.015 & 1.07E-09 & 5.00E-11 & 4.51 & m \\
46 & Pd & 1.78 & 0.02 & 1.39 & 0.06 & 4.51E-09 & 2.10E-10 & 4.65 & m \\
47 & Ag & 1.34 & 0.04 & 0.503 & 0.046 & 1.66E-09 & 1.50E-10 & 9.23 & m \\
48 & Cd & 1.84 & 0.03 & 1.59 & 0.11 & 5.48E-09 & 3.80E-10 & 6.91 & m \\
49 & In & 0.89 & 0.03 & 0.179 & 0.013 & 6.28E-10 & 4.40E-11 & 7.06 & m \\
50 & Sn & 2.21 & 0.05 & 3.69 & 0.44 & 1.34E-08 & 1.60E-09 & 12.03 & m \\
51 & Sb & 1.17 & 0.07 & 0.339 & 0.056 & 1.26E-09 & 2.10E-10 & 16.59 & m \\
52 & Te & 2.32 & 0.02 & 4.74 & 0.19 & 1.85E-08 & 7.00E-10 & 3.92 & m \\
53 & I & 1.84 & 0.15 & 1.6 & 0.64 & 6.21E-09 & 2.50E-09 & 40.24 & m \\
54 & Xe & 2.39 & 0.12 & 5.6 & 1.79 & 2.25E-08 & 7.20E-09 & 31.89 & t \\
55 & Cs & 1.2 & 0.03 & 0.367 & 0.024 & 1.49E-09 & 1.00E-10 & 6.48 & m \\
56 & Ba & 2.3 & 0.02 & 4.61 & 0.27 & 1.94E-08 & 1.13E-09 & 5.82 & m \\
57 & La & 1.31 & 0.02 & 0.469 & 0.019 & 1.99E-09 & 8.00E-11 & 4.05 & m \\
58 & Ce & 1.71 & 0.02 & 1.19 & 0.05 & 5.09E-09 & 2.10E-10 & 4.17 & m \\
59 & Pr & 0.89 & 0.02 & 0.179 & 0.008 & 7.70E-10 & 3.40E-11 & 4.4 & m \\
60 & Nd & 1.58 & 0.02 & 0.877 & 0.046 & 3.87E-09 & 2.02E-10 & 5.22 & m \\
61 & Pm & ... & ... & ... & ... & ... & ... & ... & ... \\
62 & Sm & 1.08 & 0.02 & 0.274 & 0.014 & 1.26E-09 & 7.00E-11 & 5.2 & m \\
63 & Eu & 0.66 & 0.02 & 0.104 & 0.005 & 4.84E-10 & 2.20E-11 & 4.48 & m \\
64 & Gd & 1.19 & 0.02 & 0.352 & 0.017 & 1.69E-09 & 8.00E-11 & 4.9 & m \\
65 & Tb & 0.44 & 0.02 & 0.0637 & 0.0028 & 3.10E-10 & 1.34E-11 & 4.32 & m \\
66 & Dy & 1.26 & 0.02 & 0.42 & 0.018 & 2.09E-09 & 8.90E-11 & 4.25 & m \\
67 & Ho & 0.6 & 0.02 & 0.09059 & 0.00366 & 4.57E-10 & 1.85E-11 & 4.04 & m \\
68 & Er & 1.06 & 0.02 & 0.263 & 0.011 & 1.34E-09 & 5.80E-11 & 4.32 & m \\
69 & Tm & 0.25 & 0.02 & 0.041 & 0.0019 & 2.12E-10 & 1.00E-11 & 4.57 & m \\
70 & Yb & 1.05 & 0.02 & 0.259 & 0.01 & 1.37E-09 & 5.50E-11 & 4 & m \\
71 & Lu & 0.22 & 0.02 & 0.0384 & 0.0022 & 2.05E-10 & 1.17E-11 & 5.7 & m \\
72 & Hf & 0.84 & 0.02 & 0.157 & 0.009 & 8.59E-10 & 4.97E-11 & 5.79 & m \\
73 & Ta & -0.02 & 0.02 & 0.0217 & 0.001 & 1.20E-10 & 5.60E-12 & 4.69 & m \\
74 & W & 0.8 & 0.05 & 0.144 & 0.017 & 8.10E-10 & 9.67E-11 & 11.94 & m \\
75 & Re & 0.41 & 0.02 & 0.0591 & 0.0024 & 3.36E-10 & 1.40E-11 & 4.15 & m \\
76 & Os & 1.47 & 0.01 & 0.676 & 0.019 & 3.93E-09 & 1.12E-10 & 2.86 & m \\
77 & Ir & 1.44 & 0.03 & 0.624 & 0.037 & 3.67E-09 & 2.18E-10 & 5.93 & m \\
78 & Pt & 1.73 & 0.03 & 1.22 & 0.09 & 7.27E-09 & 5.67E-10 & 7.8 & m \\
79 & Au & 0.94 & 0.04 & 0.201 & 0.018 & 1.21E-09 & 1.06E-10 & 8.8 & m \\
80 & Hg & 1.22 & 0.13 & 0.384 & 0.139 & 2.36E-09 & 8.54E-10 & 36.21 & m \\
81 & Tl & 0.9 & 0.04 & 0.181 & 0.018 & 1.13E-09 & 1.13E-10 & 9.95 & m \\
82 & Pb & 2.16 & 0.03 & 3.29 & 0.24 & 2.09E-08 & 1.52E-09 & 7.28 & m \\
83 & Bi & 0.79 & 0.03 & 0.142 & 0.009 & 9.06E-10 & 5.64E-11 & 6.22 & m \\
90 & Th & 0.27 & 0.04 & 0.0431 & 0.0047 & 3.06E-10 & 3.33E-11 & 10.89 & m \\
92 & U & 0.02 & 0.04 & 0.02382 & 0.00227 & 1.73E-10 & 1.65E-11 & 9.53 & m \\
\end{longtable}
\begin{tablenotes}
\small
\item $^{*}$Source of data: a: average of solar and meteoritic values. m: meteoritic. s: solar. t: theoretical, by other means,  see text.
\end{tablenotes}
\end{ThreePartTable}

%% file: latex_table_5.tex
%
%
\begin{ThreePartTable}
\renewcommand{\LTcapwidth}{\linewidth}
\begin{longtable}{lcccccccccccc}
\caption{\label{tab:5}Proto Solar (4.567 Ga ago) Isotopic and Elemental Compositions, Mass Fractions, and Atomic Weights*.}\\
        \toprule
          &   & & Isotope     & Elemental    &  Isotopic       &  Elemental  & Isotopic  & Atomic  & Mean  & Mean    \\
          &   & & Fractions  &  Abundance &  Abundance &  Mass        &  Mass      & Mass   &  Atomic & Atomic  \\
          &   & & of Element &                    &                      &  Fractions  &  Fractions &            &  Weight & Weight \\
          &   &  &  &  &  &  &  &  & (Proto-Solar) & (Present-Day) \\
        E & Z & A & atom\% & N(Si)$=1e6$ & $\sum \rm{Si} =1e6$ & & & AMU, Dalton & & & \\
        \midrule
        \endfirsthead
        \caption{continued.}\\
        \toprule
          &   & & Isotope     & Elemental    &  Isotopic      &  Elemental  & Isotopic  & Atomic  & Mean  & Mean    \\
          &   & & Fractions  &  Abundance &  Abundance &  Mass        &  Mass      & Mass   &  Atomic & Atomic  \\
          &   & & of Element &                   &                     &  Fractions &  Fractions &            &  Weight & Weight \\
          &   &  &  &  &  &  &  & & (Proto-Solar) & (Present-Day) \\
        E & Z & A & atom\% & N(Si)$=1e6$ & $\sum \rm{Si} =1e6$ & & & AMU, Dalton & Dalton & Dalton & \\
        \midrule
        \endhead
        \bottomrule
        \endfoot
H & 1 & 1 & 99.99803 &  & 2.29E+10 &  & 7.06E-01 & 1.007825032 &  &                             \\                
H (D) & 1 & 2 & 0.00197 &  & 4.51E+05 &  & 2.78E-05 & 2.014101778 &  & 			    \\
H & 1 &  & 100 & 2.29E+10 &  & 7.06E-01 &  &  & 1.007825 & 1.007845				    \\
He & 2 & 3 & 0.0166 &  & 3.73E+05 &  & 3.44E-05 & 3.01602932 &  & 				    \\
He & 2 & 4 & 99.9834 &  & 2.25E+09 &  & 2.75E-01 & 4.002603254 &  & 			    \\
He & 2 &  & 100 & 2.25E+09 &  & 2.75E-01 &  &  & 4.002199 & 4.002439			    \\
Li & 3 & 6 & 7.589 &  & 4.3 &  & 7.83E-10 & 6.015122885 &  & 				    \\
Li & 3 & 7 & 92.411 &  & 51.8 &  & 1.11E-08 & 7.016003428 &  & 				    \\
Li & 3 &  & 100 & 56.1 &  & 1.19E-08 &  &  & 6.940047 & 6.940047				    \\
Be & 4 & 9 & 100 & 0.657 & 0.657 & 1.81E-10 & 1.81E-10 & 9.01218291 & 9.012183 & 9.012183	    \\
B & 5 & 10 & 19.83 &  & 3.6 &  & 1.10E-09 & 10.01293696 &  & 				    \\
B & 5 & 11 & 80.17 &  & 14.5 &  & 4.89E-09 & 11.00930537 &  & 				    \\
B & 5 &  & 100 & 18.1 &  & 5.98E-09 &  &  & 10.811755 & 10.811755				    \\
C & 6 & 12 & 98.965 &  & 8.99E+06 &  & 3.30E-03 & 12 &  & 					    \\
C & 6 & 13 & 1.035 &  & 94000 &  & 3.74E-05 & 13.00335484 &  & 				    \\
C & 6 &  & 100 & 9.08E+06 &  & 3.34E-03 &  &  & 12.010389 & 12.010389			    \\
N & 7 & 14 & 99.774 &  & 2.44E+06 &  & 1.04E-03 & 14.003074 &  & 				    \\
N & 7 & 15 & 0.226 &  & 5520 &  & 2.53E-06 & 15.0001089 &  & 				    \\
N & 7 &  & 100 & 2.44E+06 &  & 1.05E-03 &  &  & 14.005325 & 14.005325			    \\
O & 8 & 16 & 99.777 &  & 1.61E+07 &  & 7.88E-03 & 15.99491462 &  & 				    \\
O & 8 & 17 & 0.035 &  & 5700 &  & 2.96E-06 & 16.99913176 &  & 				    \\
O & 8 & 18 & 0.188 &  & 30400 &  & 1.67E-05 & 17.99915961 &  & 				    \\
O & 8 &  & 100 & 1.62E+07 &  & 7.90E-03 &  &  & 15.999042 & 15.999042			    \\
F & 9 & 19 & 100 & 1278 & 1278 & 7.42E-07 & 7.42E-07 & 18.99840317 & 18.998403 & 18.998403	    \\
Ne & 10 & 20 & 93.125 &  & 3.69E+06 &  & 2.26E-03 & 19.99244018 &  & 			    \\
Ne & 10 & 21 & 0.224 &  & 8860 &  & 5.69E-06 & 20.99384668 &  & 				    \\
Ne & 10 & 22 & 6.651 &  & 2.64E+05 &  & 1.77E-04 & 21.99138512 &  & 			    \\
Ne & 10 &  & 100 & 3.96E+06 &  & 2.44E-03 &  &  & 20.127635 & 20.127635			    \\
Na & 11 & 23 & 100 & 55900 & 55900 & 3.93E-05 & 3.93E-05 & 22.98976928 & 22.989769 & 22.989769  \\
Mg & 12 & 24 & 78.992 &  & 8.18E+05 &  & 6.00E-04 & 23.9850417     \\
Mg & 12 & 25 & 10.003 &  & 1.04E+05 &  & 7.92E-05 & 24.98583691     \\
Mg & 12 & 26 & 11.005 &  & 1.14E+05 &  & 9.06E-05 & 25.98259295     \\
Mg & 12 &  & 100 & 1.04E+06 &  & 7.70E-04 &  &  & 24.304982 & 24.304982			    \\
Al & 13 & 27 & 100 & 81100 & 81100 & 6.69E-05 & 6.69E-05 & 26.98153859 & 26.981538 & 26.981538  \\
Si & 14 & 28 & 92.2297 &  & 9.22E+05 &  & 7.89E-04 & 27.97692653     \\
Si & 14 & 29 & 4.6832 &  & 46800 &  & 4.15E-05 & 28.97649467 &  & 				    \\
Si & 14 & 30 & 3.0872 &  & 30900 &  & 2.83E-05 & 29.97377017 &  & 				    \\
Si & 14 &  & 100 & 1.00E+06 &  & 8.59E-04 &  &  & 28.085384 & 28.085384			    \\
P & 15 & 31 & 100 & 8390 & 8390 & 7.95E-06 & 7.95E-06 & 30.973762 & 30.973762 & 30.973762	    \\
S & 16 & 32 & 95.04074 &  & 4.04E+05 &  & 3.95E-04 & 31.97207117     \\
S & 16 & 33 & 0.74869 &  & 3180 &  & 3.21E-06 & 32.97145569 &  & 				    \\
S & 16 & 34 & 4.19599 &  & 17800 &  & 1.85E-05 & 33.9678669 &  & 				    \\
S & 16 & 36 & 0.01458 &  & 62 &  & 6.82E-08 & 35.96708076 &  & 				    \\
S & 16 &  & 100 & 4.25E+05 &  & 4.17E-04 &  &  & 32.063879 & 32.063879			    \\
Cl & 17 & 35 & 75.7647 &  & 4030 &  & 4.31E-06 & 34.96885268 &  & 				    \\
Cl & 17 & 37 & 24.2353 &  & 1290 &  & 1.46E-06 & 36.96590259 &  & 				    \\
Cl & 17 &  & 100 & 5320 &  & 5.77E-06 &  &  & 35.452844 & 35.452844				    \\
Ar & 18 & 36 & 84.596 &  & 75100 &  & 8.26E-05 & 35.96754511 &  & 				    \\
Ar & 18 & 38 & 15.38 &  & 13600 &  & 1.58E-05 & 37.96273234 &  & 				    \\
Ar$*$ & 18 & 40 & 0.024 &  & 21 &  & 2.57E-08 & 39.96238312 &  & 				    \\
Ar & 18 &  & 100 & 88800 &  & 9.84E-05 &  &  & 36.275378 & 36.275357			    \\
K & 19 & 39 & 93.132 &  & 3410 &  & 4.06E-06 & 38.96370649 &  & 				    \\
K$*$ & 19 & 40 & 0.147 &  & 5 &  & 6.11E-09 & 39.96399848 &  & 				    \\
K & 19 & 41 & 6.721 &  & 246 &  & 3.08E-07 & 40.96182526 &  & 				    \\
K & 19 &  & 100 & 3660 &  & 4.38E-06 &  &  & 39.098302 & 39.099469				    \\
Ca$*$ & 20 & 40 & 96.941 &  & 58500 &  & 7.15E-05 & 39.96259086 &  & 				    \\
Ca & 20 & 42 & 0.647 &  & 391 &  & 5.02E-07 & 41.95861801 &  & 				    \\
Ca & 20 & 43 & 0.135 &  & 82 &  & 1.08E-07 & 42.95876667 &  & 				    \\
Ca & 20 & 44 & 2.086 &  & 1260 &  & 1.69E-06 & 43.95548173 &  & 				    \\
Ca & 20 & 46 & 0.004 &  & 2 &  & 2.81E-09 & 45.9536926 &  & 				    \\
Ca & 20 & 48 & 0.187 &  & 113 &  & 1.66E-07 & 47.9525343 &  & 				    \\
Ca & 20 &  & 100 & 60400 &  & 7.40E-05 &  &  & 40.078022 & 40.078022			    \\
Sc & 21 & 45 & 100 & 33.7 & 3.37E+01 & 4.63E-08 & 4.63E-08 & 44.9559119 & 44.955907 & 44.955907 \\
Ti & 22 & 46 & 8.249 &  & 200 &  & 2.81E-07 & 45.95262889 &  & 				    \\
Ti & 22 & 47 & 7.437 &  & 181 &  & 2.60E-07 & 46.95176293 &  & 				    \\
Ti & 22 & 48 & 73.72 &  & 1790 &  & 2.63E-06 & 47.94794631 &  & 				    \\
Ti & 22 & 49 & 5.409 &  & 131 &  & 1.96E-07 & 48.94786998 &  & 				    \\
Ti & 22 & 50 & 5.185 &  & 126 &  & 1.93E-07 & 49.94479117 &  & 				    \\
Ti & 22 &  & 100 & 2430 &  & 3.55E-06 &  &  & 47.866883 & 47.866883				    \\
V & 23 & 50 & 0.25 &  & 0.7 &  & 1.05E-09 & 49.9471585 &  & 				    \\
V & 23 & 51 & 99.75 &  & 273.6 &  & 4.26E-07 & 50.9439595 &  & 				    \\
V & 23 &  & 100 & 274 &  & 4.27E-07 &  &  & 50.941469 & 50.941469				    \\
Cr & 24 & 50 & 4.345 &  & 575 &  & 8.78E-07 & 49.9460442 &  & 				    \\
Cr & 24 & 52 & 83.79 &  & 11100 &  & 1.76E-05 & 51.94050751 &  & 				    \\
Cr & 24 & 53 & 9.501 &  & 1260 &  & 2.04E-06 & 52.94064943 &  & 				    \\
Cr & 24 & 54 & 2.365 &  & 313 &  & 5.16E-07 & 53.93888045 &  & 				    \\
Cr & 24 &  & 100 & 13200 &  & 2.11E-05 &  &  & 51.996116 & 51.996116			    \\
Mn & 25 & 55 & 100 & 9270 & 9.27E+03 & 1.56E-05 & 1.56E-05 & 54.93804512 & 54.938043 & 54.938043\\
Fe & 26 & 54 & 5.845 &  & 51100 &  & 8.43E-05 & 53.93961046 &  & 				    \\
Fe & 26 & 56 & 91.754 &  & 8.02E+05 &  & 1.37E-03 & 55.93493745     \\
Fe & 26 & 57 & 2.119 &  & 18500 &  & 3.22E-05 & 56.93539427 &  & 				    \\
Fe & 26 & 58 & 0.282 &  & 2460 &  & 4.36E-06 & 57.93327558 &  & 				    \\
Fe & 26 &  & 100 & 873500 &  & 1.49E-03 &  &  & 55.845143 & 55.845143			    \\
Co & 27 & 59 & 100 & 2290 & 2.29E+03 & 4.07E-06 & 4.07E-06 & 58.93319506 & 58.933194 & 58.933194\\
Ni & 28 & 58 & 68.077 &  & 34100 &  & 6.04E-05 & 57.9353435 &  & 				    \\
Ni & 28 & 60 & 26.223 &  & 13100 &  & 2.40E-05 & 59.93078635 &  & 				    \\
Ni & 28 & 61 & 1.14 &  & 570 &  & 1.06E-06 & 60.93105603 &  & 				    \\
Ni & 28 & 62 & 3.635 &  & 1820 &  & 3.45E-06 & 61.92834511 &  & 				    \\
Ni & 28 & 64 & 0.926 &  & 463 &  & 9.05E-07 & 63.92796594 &  & 				    \\
Ni & 28 &  & 100 & 50030 &  & 8.99E-05 &  &  & 58.69335 & 58.69335				    \\
Cu & 29 & 63 & 69.174 &  & 381 &  & 7.34E-07 & 62.92959751 &  & 				    \\
Cu & 29 & 65 & 30.826 &  & 170 &  & 3.37E-07 & 64.92778945 &  & 				    \\
Cu & 29 &  & 100 & 551 &  & 1.07E-06 &  &  & 63.54556 & 63.54556				    \\
Zn & 30 & 64 & 49.1704 &  & 614 &  & 1.20E-06 & 63.92914224 &  & 				    \\
Zn & 30 & 66 & 27.7306 &  & 346 &  & 6.98E-07 & 65.92603345 &  & 				    \\
Zn & 30 & 67 & 4.0401 &  & 50 &  & 1.03E-07 & 66.92712739 &  & 				    \\
Zn & 30 & 68 & 18.4483 &  & 230 &  & 4.78E-07 & 67.9248442 &  & 				    \\
Zn & 30 & 70 & 0.6106 &  & 8 &  & 1.63E-08 & 69.9253193 &  & 				    \\
Zn & 30 &  & 100 & 1250 &  & 2.50E-06 &  &  & 65.377765 & 65.377765				    \\
Ga & 31 & 69 & 60.108 &  & 21.6 &  & 4.56E-08 & 68.9255735 &  & 				    \\
Ga & 31 & 71 & 39.892 &  & 14.4 &  & 3.11E-08 & 70.9247026 &  & 				    \\
Ga & 31 &  & 100 & 36 &  & 7.67E-08 &  &  & 69.723068 & 69.723068				    \\
Ge & 32 & 70 & 20.526 &  & 24.9 &  & 5.33E-08 & 69.9242474 &  & 				    \\
Ge & 32 & 72 & 27.446 &  & 33.3 &  & 7.33E-08 & 71.9220758 &  & 				    \\
Ge & 32 & 73 & 7.76 &  & 9.4 &  & 2.10E-08 & 72.9234589 &  & 				    \\
Ge & 32 & 74 & 36.523 &  & 44.3 &  & 1.00E-07 & 73.92117777 &  & 				    \\
Ge & 32 & 76 & 7.745 &  & 9.4 &  & 2.18E-08 & 75.92140273 &  & 				    \\
Ge & 32 &  & 100 & 121 &  & 2.70E-07 &  &  & 72.629589 & 72.629589				    \\
As & 33 & 75 & 100 & 6.15 & 6.15 & 1.41E-08 & 1.41E-08 & 74.9215965 & 74.921595 & 74.921595	    \\
Se & 34 & 74 & 0.863 &  & 0.6 &  & 1.40E-09 & 73.92247594 &  & 				    \\
Se & 34 & 76 & 9.22 &  & 6.6 &  & 1.53E-08 & 75.91921372 &  & 				    \\
Se & 34 & 77 & 7.594 &  & 5.4 &  & 1.28E-08 & 76.919914 &  & 				    \\
Se & 34 & 78 & 23.685 &  & 16.9 &  & 4.04E-08 & 77.9173091 &  & 				    \\
Se & 34 & 80 & 49.813 &  & 35.6 &  & 8.71E-08 & 79.9165213 &  & 				    \\
Se$^{\wedge}$ & 34 & 82 & 8.825 &  & 6.3 &  & 1.58E-08 & 81.9166994 &  & 				    \\
Se & 34 &  & 100 & 71.5 &  & 1.73E-07 &  &  & 78.971681 & 78.971681				    \\
Br & 35 & 79 & 50.686 &  & 6.29 &  & 1.52E-08 & 78.9183371 &  & 				    \\
Br & 35 & 81 & 49.314 &  & 6.12 &  & 1.52E-08 & 80.9162906 &  & 				    \\
Br & 35 &  & 100 & 12.4 &  & 3.03E-08 &  &  & 79.903607 & 79.903607				    \\
Kr & 36 & 78 & 0.36526667 &  & 0.21 &  & 4.99E-10 & 77.92036486     \\
Kr & 36 & 80 & 2.34407892 &  & 1.34 &  & 3.28E-09 & 79.91637915     \\
Kr & 36 & 82 & 11.6862576 &  & 6.7 &  & 1.68E-08 & 81.91348282     \\
Kr & 36 & 83 & 11.5724673 &  & 6.63 &  & 1.68E-08 & 82.9141271     \\
Kr & 36 & 84 & 56.8951195 &  & 32.6 &  & 8.37E-08 & 83.91149717     \\
Kr & 36 & 86 & 17.13681 &  & 9.82 &  & 2.58E-08 & 85.91061067 &  & 				    \\
Kr & 36 &  & 100 & 57.3 &  & 1.47E-07 &  &  & 83.789635 & 83.789635				    \\
Rb & 37 & 85 & 70.844 &  & 5.02 &  & 1.31E-08 & 84.91178974 &  & 				    \\
Rb$*$ & 37 & 87 & 29.156 &  & 2.07 &  & 5.50E-09 & 86.90918054 &  & 				    \\
Rb & 37 &  & 100 & 7.09 &  & 1.86E-08 &  &  & 85.467755 & 85.494154				    \\
Sr & 38 & 84 & 0.558 &  & 0.13 &  & 3.44E-10 & 83.9134203 &  & 				    \\
Sr & 38 & 86 & 9.871 &  & 2.37 &  & 6.23E-09 & 85.9092602 &  & 				    \\
Sr & 38 & 87 & 6.898 &  & 1.66 &  & 4.40E-09 & 86.9088771 &  & 				    \\
Sr & 38 & 88 & 82.672 &  & 19.8 &  & 5.34E-08 & 87.9056122 &  & 				    \\
Sr & 38 &  & 100 & 24 &  & 6.43E-08 &  &  & 87.613691 & 87.617504				    \\
Y & 39 & 89 & 100 & 4.49 & 4.49E+00 & 1.22E-08 & 1.22E-08 & 88.9058483 & 88.905838 & 88.905838  \\
Zr & 40 & 90 & 51.49 &  & 5.41 &  & 1.49E-08 & 89.9047044 &  & 				    \\
Zr & 40 & 91 & 11.218 &  & 1.18 &  & 3.28E-09 & 90.9056458 &  & 				    \\
Zr & 40 & 92 & 17.148 &  & 1.8 &  & 5.06E-09 & 91.9050408 &  & 				    \\
Zr & 40 & 94 & 17.359 &  & 1.82 &  & 5.24E-09 & 93.9063152 &  & 				    \\
Zr$^{\wedge}$ & 40 & 96 & 2.785 &  & 0.29 &  & 8.58E-10 & 95.9082734 &  & 				    \\
Zr & 40 &  & 100 & 10.5 &  & 2.93E-08 &  &  & 91.221842 & 91.221842				    \\
Nb & 41 & 93 & 100 & 0.766 & 0.766 & 2.18E-09 & 2.18E-09 & 92.9063781 & 92.906373 & 92.906373   \\
Mo & 42 & 92 & 14.649904 &  & 0.38 &  & 1.07E-09 & 91.90680811     \\
Mo & 42 & 94 & 9.1877391 &  & 0.238 &  & 6.85E-10 & 93.9050856     \\
Mo & 42 & 95 & 15.8737718 &  & 0.412 &  & 1.20E-09 & 94.9058394     \\
Mo & 42 & 96 & 16.6738099 &  & 0.433 &  & 1.27E-09 & 95.90467712     \\
Mo & 42 & 97 & 9.58299602 &  & 0.249 &  & 7.37E-10 & 96.9060196     \\
Mo & 42 & 98 & 24.2868708 &  & 0.63 &  & 1.89E-09 & 97.9054058     \\
Mo$^{\wedge}$ & 42 & 100 & 9.74490849 &  & 0.253 &  & 7.73E-10 & 99.9074724     \\
Mo & 42 &  & 100 & 2.6 &  & 7.62E-09 &  &  & 95.948662 & 95.948662				    \\
Ru & 44 & 96 & 5.54 &  & 0.098 &  & 2.87E-10 & 95.9075939 &  & 				    \\
Ru & 44 & 98 & 1.87 &  & 0.033 &  & 9.89E-11 & 97.9052876 &  & 				    \\
Ru & 44 & 99 & 12.76 &  & 0.226 &  & 6.82E-10 & 98.9059393 &  & 				    \\
Ru & 44 & 100 & 12.6 &  & 0.223 &  & 6.81E-10 & 99.9042195 &  & 				    \\
Ru & 44 & 101 & 17.06 &  & 0.302 &  & 9.31E-10 & 100.9055821 &  & 				    \\
Ru & 44 & 102 & 31.55 &  & 0.558 &  & 1.74E-09 & 101.9043493 &  & 				    \\
Ru & 44 & 104 & 18.62 &  & 0.329 &  & 1.05E-09 & 103.9054326 &  & 				    \\
Ru & 44 &  & 100 & 1.77 &  & 5.46E-09 &  &  & 101.06498 & 101.06498				    \\
Rh & 45 & 103 & 100 & 0.341 & 0.341 & 1.07E-09 & 1.07E-09 & 102.9055043 & 102.90549 & 102.90549 \\
Pd & 46 & 102 & 1.02 &  & 0.014 &  & 4.40E-11 & 101.9056286 &  & 				    \\
Pd & 46 & 104 & 11.14 &  & 0.154 &  & 4.90E-10 & 103.9040359 &  & 				    \\
Pd & 46 & 105 & 22.33 &  & 0.309 &  & 9.93E-10 & 104.9050847 &  & 				    \\
Pd & 46 & 106 & 27.33 &  & 0.379 &  & 1.23E-09 & 105.9034808 &  & 				    \\
Pd & 46 & 108 & 26.46 &  & 0.367 &  & 1.21E-09 & 107.9038907 &  & 				    \\
Pd & 46 & 110 & 11.72 &  & 0.162 &  & 5.46E-10 & 109.9051703 &  & 				    \\
Pd & 46 &  & 100 & 1.39 &  & 4.51E-09 &  &  & 106.41533 & 106.41533				    \\
Ag & 47 & 107 & 51.8392 &  & 0.261 &  & 8.52E-10 & 106.9050965     \\
Ag & 47 & 109 & 48.1608 &  & 0.242 &  & 8.06E-10 & 108.9047523     \\
Ag & 47 &  & 100 & 0.503 &  & 1.66E-09 &  &  & 107.86815 & 107.86815			    \\
Cd & 48 & 106 & 1.249 &  & 0.02 &  & 6.45E-11 & 105.9064602 &  & 				    \\
Cd & 48 & 108 & 0.89 &  & 0.014 &  & 4.68E-11 & 107.9041824 &  & 				    \\
Cd & 48 & 110 & 12.485 &  & 0.199 &  & 6.69E-10 & 109.9030035 &  & 				    \\
Cd & 48 & 111 & 12.804 &  & 0.204 &  & 6.92E-10 & 110.9041781 &  & 				    \\
Cd & 48 & 112 & 24.117 &  & 0.385 &  & 1.32E-09 & 111.9027578 &  & 				    \\
Cd$^{\wedge}$ & 48 & 113 & 12.225 &  & 0.195 &  & 6.73E-10 & 112.9044026     \\
Cd & 48 & 114 & 28.729 &  & 0.458 &  & 1.60E-09 & 113.9033595 &  & 				    \\
Cd$^{\wedge}$ & 48 & 116 & 7.501 &  & 0.12 &  & 4.24E-10 & 115.9047632 &  & 				    \\
Cd & 48 &  & 100 & 1.59 &  & 5.48E-09 &  &  & 112.41215 & 112.41215				    \\
In & 49 & 113 & 4.281 &  & 0.008 &  & 2.64E-11 & 112.9040574 &  & 				    \\
In$^{\wedge}$ & 49 & 115 & 95.719 &  & 0.171 &  & 6.02E-10 & 114.9038788     \\
In & 49 &  & 100 & 0.179 &  & 6.28E-10 &  &  & 114.81827 & 114.81827			    \\
Sn & 50 & 112 & 0.971 &  & 0.036 &  & 1.23E-10 & 111.9048218 &  & 				    \\
Sn & 50 & 114 & 0.659 &  & 0.024 &  & 8.47E-11 & 113.9027788 &  & 				    \\
Sn & 50 & 115 & 0.339 &  & 0.013 &  & 4.40E-11 & 114.9033424 &  & 				    \\
Sn & 50 & 116 & 14.536 &  & 0.537 &  & 1.90E-09 & 115.9017405 &  & 				    \\
Sn & 50 & 117 & 7.676 &  & 0.283 &  & 1.01E-09 & 116.9029516 &  & 				    \\
Sn & 50 & 118 & 24.223 &  & 0.894 &  & 3.22E-09 & 117.9016031 &  & 				    \\
Sn & 50 & 119 & 8.59 &  & 0.317 &  & 1.15E-09 & 118.9033076 &  & 				    \\
Sn & 50 & 120 & 32.593 &  & 1.203 &  & 4.41E-09 & 119.9022002 &  & 				    \\
Sn & 50 & 122 & 4.629 &  & 0.171 &  & 6.37E-10 & 121.9034391 &  & 				    \\
Sn & 50 & 124 & 5.789 &  & 0.214 &  & 8.10E-10 & 123.9052761 &  & 				    \\
Sn & 50 &  & 100 & 3.69 &  & 1.34E-08 &  &  & 118.71035 & 118.71035				    \\
Sb & 51 & 121 & 57.213 &  & 0.194 &  & 7.18E-10 & 120.9038157 &  & 				    \\
Sb & 51 & 123 & 42.787 &  & 0.145 &  & 5.46E-10 & 122.904214 &  & 				    \\
Sb & 51 &  & 100 & 0.339 &  & 1.26E-09 &  &  & 121.75972 & 121.75972			    \\
Te & 52 & 120 & 0.096 &  & 0.005 &  & 1.67E-11 & 119.9040452 &  & 				    \\
Te & 52 & 122 & 2.603 &  & 0.124 &  & 4.61E-10 & 121.9030439 &  & 				    \\
Te$^{\wedge}$ & 52 & 123 & 0.908 &  & 0.043 &  & 1.62E-10 & 122.9042701 &  & 				    \\
Te & 52 & 124 & 4.816 &  & 0.229 &  & 8.66E-10 & 123.9028176 &  & 				    \\
Te & 52 & 125 & 7.139 &  & 0.339 &  & 1.29E-09 & 124.9044307 &  & 				    \\
Te & 52 & 126 & 18.952 &  & 0.899 &  & 3.46E-09 & 125.9033117 &  & 				    \\
Te$^{\wedge}$ & 52 & 128 & 31.687 &  & 1.504 &  & 5.88E-09 & 127.9044621     \\
Te$^{\wedge}$ & 52 & 130 & 33.799 &  & 1.604 &  & 6.37E-09 & 129.9062228     \\
Te & 52 &  & 100 & 4.74 &  & 1.85E-08 &  &  & 127.58559 & 127.58559				    \\
I & 53 & 127 & 100 & 1.59 & 1.59E+00 & 6.17E-09 & 6.17E-09 & 126.9044728 & 126.90447 & 126.90447 \\
Xe$^{\wedge}$ & 54 & 124 & 0.129 &  & 0.007 &  & 2.75E-11 & 123.905893 &  & 				    \\
Xe & 54 & 126 & 0.11 &  & 0.006 &  & 2.37E-11 & 125.9042912 &  & 				    \\
Xe & 54 & 128 & 2.22 &  & 0.124 &  & 4.86E-10 & 127.9035313 &  & 				    \\
Xe & 54 & 129 & 27.428 &  & 1.536 &  & 6.06E-09 & 128.9047809 &  & 				    \\
Xe & 54 & 130 & 4.349 &  & 0.244 &  & 9.68E-10 & 129.9035094 &  & 				    \\
Xe & 54 & 131 & 21.763 &  & 1.219 &  & 4.88E-09 & 130.9050524 &  & 				    \\
Xe & 54 & 132 & 26.36 &  & 1.476 &  & 5.96E-09 & 131.9041551 &  & 				    \\
Xe & 54 & 134 & 9.73 &  & 0.545 &  & 2.23E-09 & 133.9053945 &  & 				    \\
Xe$^{\wedge}$ & 54 & 136 & 7.911 &  & 0.443 &  & 1.84E-09 & 135.9072145 &  & 				    \\
Xe & 54 &  & 100 & 5.6 &  & 2.25E-08 &  &  & 131.1827 & 131.18269				    \\
Cs & 55 & 133 & 100 & 0.367 & 3.67E-01 & 1.49E-09 & 1.49E-09 & 132.905452 & 132.90545 & 132.90545 \\
Ba$^{\wedge}$ & 56 & 130 & 0.1058 &  & 4.88E-03 &  & 1.94E-11 & 129.9063215     \\
Ba & 56 & 132 & 0.1012 &  & 4.67E-03 &  & 1.88E-11 & 131.9050613     \\
Ba & 56 & 134 & 2.417 &  & 1.12E-01 &  & 4.56E-10 & 133.9045084     \\
Ba & 56 & 135 & 6.592 &  & 3.04E-01 &  & 1.25E-09 & 134.9056886     \\
Ba & 56 & 136 & 7.853 &  & 3.62E-01 &  & 1.51E-09 & 135.904576     \\
Ba & 56 & 137 & 11.232 &  & 5.18E-01 &  & 2.17E-09 & 136.9058274     \\
Ba & 56 & 138 & 71.699 &  & 3.31E+00 &  & 1.39E-08 & 137.9052473     \\
Ba & 56 &  & 100 & 4.61 &  & 1.94E-08 &  &  & 137.32692 & 137.32692				    \\
La$*$ & 57 & 138 & 0.0916 &  & 4.29E-04 &  & 1.81E-12 & 137.907112     \\
La & 57 & 139 & 99.9084 &  & 4.68E-01 &  & 1.99E-09 & 138.9063533     \\
La & 57 &  & 100 & 0.469 &  & 1.99E-09 &  &  & 138.90548 & 138.90545			    \\
Ce & 58 & 136 & 0.186 &  & 2.21E-03 &  & 9.18E-12 & 135.9071295     \\
Ce$^{\wedge}$ & 58 & 138 & 0.25 &  & 2.97E-03 &  & 1.25E-11 & 137.905991     \\
Ce & 58 & 140 & 88.45 &  & 1.05E+00 &  & 4.49E-09 & 139.9054387     \\
Ce$^{\wedge}$ & 58 & 142 & 11.114 &  & 1.32E-01 &  & 5.73E-10 & 141.9092442     \\
Ce & 58 &  & 100 & 1.19 &  & 5.09E-09 &  &  & 140.1157 & 140.11572				    \\
Pr & 59 & 141 & 100 & 0.179 & 1.79E-01 & 7.70E-10 & 7.70E-10 & 140.9076525 & 140.90766 & 140.90766 \\
Nd & 60 & 142 & 27.045 &  & 0.237 &  & 1.03E-09 & 141.9077233 &  & 				    \\
Nd$*$ & 60 & 143 & 12.023 &  & 0.105 &  & 4.61E-10 & 142.9098143     \\
Nd$^{\wedge}$ & 60 & 144 & 23.729 &  & 0.208 &  & 9.16E-10 & 143.9100873     \\
Nd & 60 & 145 & 8.763 &  & 0.077 &  & 3.41E-10 & 144.9125736 &  & 				    \\
Nd & 60 & 146 & 17.13 &  & 0.15 &  & 6.71E-10 & 145.9131169 &  & 				    \\
Nd & 60 & 148 & 5.716 &  & 0.05 &  & 2.27E-10 & 147.9168933 &  & 				    \\
Nd$^{\wedge}$ & 60 & 150 & 5.596 &  & 0.049 &  & 2.25E-10 & 149.9208949 &  & 				    \\
Nd & 60 &  & 100 & 0.877 &  & 3.87E-09 &  &  & 144.24276 & 144.24465			    \\
Sm & 62 & 144 & 3.083 &  & 0.0084 &  & 3.70E-11 & 143.9120046 &  & 				    \\
Sm$*$ & 62 & 147 & 15.017 &  & 0 &  & 0.00E+00 & 146.9148979 &  & 				    \\
Sm$^{\wedge}$ & 62 & 148 & 11.254 &  & 0.0422 &  & 1.91E-10 & 147.9148227     \\
Sm & 62 & 149 & 13.83 &  & 0.0307 &  & 1.40E-10 & 148.9171847 &  & 				    \\
Sm & 62 & 150 & 7.351 &  & 0.0377 &  & 1.73E-10 & 149.9172755 &  & 				    \\
Sm & 62 & 152 & 26.735 &  & 0.0201 &  & 9.33E-11 & 151.9197324     \\
Sm & 62 & 154 & 22.73 &  & 0.0729 &  & 3.43E-10 & 153.9222093 &  & 				    \\
Sm & 62 &  & 100 & 0.062 &  & 9.77E-10 &  &  & 150.365 & 150.36328				    \\
Eu$^{\wedge}$ & 63 & 151 & 47.81 &  & 0.0498 &  & 2.30E-10 & 150.9198502     \\
Eu & 63 & 153 & 52.19 &  & 0.0543 &  & 2.54E-10 & 152.9212303 &  & 				    \\
Eu & 63 &  & 100 & 0.1041 & 0.1041 & 4.84E-10 &  &  & 151.96438 & 151.96438			    \\
Gd$^{\wedge}$ & 64 & 152 & 0.2029 &  & 0.00071 &  & 3.32E-12 & 151.9197922     \\
Gd & 64 & 154 & 2.1809 &  & 0.00768 &  & 3.62E-11 & 153.9208693     \\
Gd & 64 & 155 & 14.7998 &  & 0.05213 &  & 2.47E-10 & 154.9226276     \\
Gd & 64 & 156 & 20.4664 &  & 0.07209 &  & 3.44E-10 & 155.9221287     \\
Gd & 64 & 157 & 15.6518 &  & 0.05513 &  & 2.65E-10 & 156.9239647     \\
Gd & 64 & 158 & 24.8347 &  & 0.08747 &  & 4.23E-10 & 157.9241101     \\
Gd & 64 & 160 & 21.8635 &  & 0.07701 &  & 3.77E-10 & 159.9270585     \\
Gd & 64 &  & 100 & 0.352 &  & 1.69E-09 &  &  & 157.25205 & 157.25205			    \\
Tb & 65 & 159 & 100 & 0.0637 & 6.37E-02 & 3.10E-10 & 3.10E-10 & 158.9253468 & 158.92535 & 158.92535 \\
Dy & 66 & 156 & 0.0539 &  & 0.0002 &  & 1.08E-12 & 155.9242829     \\
Dy & 66 & 158 & 0.0946 &  & 0.0004 &  & 1.92E-12 & 157.9244096     \\
Dy & 66 & 160 & 2.3288 &  & 0.0098 &  & 4.78E-11 & 159.9251975     \\
Dy & 66 & 161 & 18.8887 &  & 0.0793 &  & 3.90E-10 & 160.9269334     \\
Dy & 66 & 162 & 25.4791 &  & 0.1069 &  & 5.30E-10 & 161.9267984     \\
Dy & 66 & 163 & 24.8954 &  & 0.1045 &  & 5.21E-10 & 162.9287312     \\
Dy & 66 & 164 & 28.2596 &  & 0.1186 &  & 5.95E-10 & 163.9291748     \\
Dy & 66 &  & 100 & 0.42 &  & 2.09E-09 &  &  & 162.49977 & 162.49977				    \\
Ho & 67 & 165 & 100 & 0.0906 & 0.0906 & 4.57E-10 & 4.57E-10 & 164.9303221 & 164.93033 & 164.93033 \\
Er & 68 & 162 & 0.139 &  & 0.0004 &  & 1.81E-12 & 161.9287799 &  & 				    \\
Er & 68 & 164 & 1.601 &  & 0.004 &  & 2.11E-11 & 163.9292065 &  & 				    \\
Er & 68 & 166 & 33.503 &  & 0.088 &  & 4.47E-10 & 165.9302931 &  & 				    \\
Er & 68 & 167 & 22.869 &  & 0.06 &  & 3.07E-10 & 166.9320482 &  & 				    \\
Er & 68 & 168 & 26.978 &  & 0.071 &  & 3.64E-10 & 167.9323702 &  & 				    \\
Er & 68 & 170 & 14.91 &  & 0.039 &  & 2.04E-10 & 169.9354643 &  & 				    \\
Er & 68 &  & 100 & 0.263 &  & 1.34E-09 &  &  & 167.25908 & 167.25908			    \\
Tm & 69 & 169 & 100 & 0.041 & 0.041 & 2.12E-10 & 2.12E-10 & 168.9342133 & 168.93422 & 168.93422 \\
Yb & 70 & 168 & 0.123 &  & 0.0003 &  & 1.64E-12 & 167.9338869 &  & 				    \\
Yb & 70 & 170 & 2.982 &  & 0.008 &  & 4.01E-11 & 169.9347618 &  & 				    \\
Yb & 70 & 171 & 14.086 &  & 0.036 &  & 1.91E-10 & 170.9363258 &  & 				    \\
Yb & 70 & 172 & 21.686 &  & 0.056 &  & 2.95E-10 & 171.9363815 &  & 				    \\
Yb & 70 & 173 & 16.103 &  & 0.042 &  & 2.20E-10 & 172.9382108 &  & 				    \\
Yb & 70 & 174 & 32.025 &  & 0.083 &  & 4.41E-10 & 173.9388621 &  & 				    \\
Yb & 70 & 176 & 12.995 &  & 0.034 &  & 1.81E-10 & 175.9425717 &  & 				    \\
Yb & 70 &  & 100 & 0.259 &  & 1.37E-09 &  &  & 173.05447 & 173.05447			    \\
Lu & 71 & 175 & 97.18 &  & 0.0373 &  & 2.00E-10 & 174.9407712 &  & 				    \\
Lu$*$ & 71 & 176 & 2.82 &  & 0.0011 &  & 5.83E-12 & 175.9426867 &  & 				    \\
Lu & 71 &  & 100 & 0.0384 &  & 2.05E-10 &  &  & 174.96681 & 174.96906			    \\
Hf$^{\wedge}$ & 72 & 174 & 0.16 &  & 0.0003 &  & 1.35E-12 & 173.9400462 &  & 				    \\
Hf & 72 & 176 & 5.2 &  & 0.008 &  & 4.41E-11 & 175.9414091 &  & 				    \\
Hf & 72 & 177 & 18.6 &  & 0.029 &  & 1.59E-10 & 176.9432224 &  & 				    \\
Hf & 72 & 178 & 27.3 &  & 0.043 &  & 2.34E-10 & 177.9437004 &  & 				    \\
Hf & 72 & 179 & 13.63 &  & 0.021 &  & 1.17E-10 & 178.945817 &  & 				    \\
Hf & 72 & 180 & 35.11 &  & 0.055 &  & 3.04E-10 & 179.9465512 &  & 				    \\
Hf & 72 &  & 100 & 0.157 &  & 8.59E-10 &  &  & 178.48515 & 178.48658			    \\
Ta$*$ & 73 & 180 & 0.01201 &  & 0.000003 &  & 1.44E-14 & 179.9474648     \\
Ta & 73 & 181 & 99.98799 &  & 0.0217 &  & 1.20E-10 & 180.9479958     \\
Ta & 73 &  & 100 & 0.0217 &  & 1.20E-10 &  &  & 180.94788 & 180.94788			    \\
W$^{\wedge}$ & 74 & 180 & 0.1198 &  & 0.0002 &  & 9.49E-13 & 179.9467091     \\
W & 74 & 182 & 26.4985 &  & 0.038 &  & 2.12E-10 & 181.9482042 &  & 				    \\
W & 74 & 183 & 14.3136 &  & 0.021 &  & 1.15E-10 & 182.9502223 &  & 				    \\
W & 74 & 184 & 30.6422 &  & 0.044 &  & 2.48E-10 & 183.9509312 &  & 				    \\
W & 74 & 186 & 28.4259 &  & 0.041 &  & 2.33E-10 & 185.9543641 &  & 				    \\
W & 74 &  & 100 & 0.144 &  & 8.10E-10 &  &  & 183.8417 & 183.8417				    \\
Re & 75 & 185 & 35.6616 &  & 0.0211 &  & 1.19E-10 & 184.9529549     \\
Re$*$ & 75 & 187 & 64.3384 &  & 0.038 &  & 2.17E-10 & 186.9557531     \\
Re & 75 &  & 100 & 0.0591 &  & 3.36E-10 &  &  & 186.20675 & 186.24152			    \\
Os$^{\wedge}$ & 76 & 184 & 0.0198 &  & 0.0001 &  & 7.55E-13 & 183.9524891     \\
Os$^{\wedge}$ & 76 & 186 & 1.5973 &  & 0.011 &  & 6.14E-11 & 185.9538382     \\
Os & 76 & 187 & 1.2817 &  & 0.009 &  & 4.96E-11 & 186.9557505 &  & 				    \\
Os & 76 & 188 & 13.3269 &  & 0.09 &  & 5.18E-10 & 187.9558382 &  & 				    \\
Os & 76 & 189 & 16.2549 &  & 0.11 &  & 6.35E-10 & 188.9581475 &  & 				    \\
Os & 76 & 190 & 26.4368 &  & 0.179 &  & 1.04E-09 & 189.9584471     \\
Os & 76 & 192 & 41.0827 &  & 0.278 &  & 1.63E-09 & 191.9614807     \\
Os & 76 &  & 100 & 0.676 &  & 3.94E-09 &  &  & 190.23494 & 190.24822			    \\
Ir & 77 & 191 & 37.272 &  & 0.232 &  & 1.36E-09 & 190.9605941 &  & 				    \\
Ir & 77 & 193 & 62.728 &  & 0.391 &  & 2.31E-09 & 192.9629264 &  & 				    \\
Ir & 77 &  & 100 & 0.624 &  & 3.67E-09 &  &  & 192.21661 & 192.21661			    \\
Pt$*$ & 78 & 190 & 0.013 &  & 0.0002 &  & 9.18E-13 & 189.9599321     \\
Pt & 78 & 192 & 0.7938 &  & 0.01 &  & 5.68E-11 & 191.961038 &  & 				    \\
Pt & 78 & 194 & 32.8078 &  & 0.4 &  & 2.37E-09 & 193.962679 &  & 				    \\
Pt & 78 & 195 & 33.7871 &  & 0.412 &  & 2.45E-09 & 194.9647901     \\
Pt & 78 & 196 & 25.2902 &  & 0.308 &  & 1.85E-09 & 195.9649515     \\
Pt & 78 & 198 & 7.3083 &  & 0.089 &  & 5.39E-10 & 197.967891 &  & 				    \\
Pt & 78 &  & 100 & 1.218 &  & 7.27E-09 &  &  & 195.08395 & 195.08395			    \\
Au & 79 & 197 & 100 & 0.201 & 0.201 & 1.21E-09 & 1.21E-09 & 196.9665687 & 196.96657 & 196.96657 \\
Hg & 80 & 196 & 0.16 &  & 0.001 &  & 3.57E-12 & 195.9658326 &  & 				    \\
Hg & 80 & 198 & 10.04 &  & 0.039 &  & 2.34E-10 & 197.9667689 &  & 				    \\
Hg & 80 & 199 & 16.94 &  & 0.065 &  & 3.96E-10 & 198.9682804 &  & 				    \\
Hg & 80 & 200 & 23.14 &  & 0.089 &  & 5.44E-10 & 199.968326 &  & 				    \\
Hg & 80 & 201 & 13.17 &  & 0.051 &  & 3.11E-10 & 200.9703022 &  & 				    \\
Hg & 80 & 202 & 29.74 &  & 0.114 &  & 7.06E-10 & 201.970643 &  & 				    \\
Hg & 80 & 204 & 6.82 &  & 0.026 &  & 1.64E-10 & 203.9734941 &  & 				    \\
Hg & 80 &  & 100 & 0.384 &  & 2.36E-09 &  &  & 200.5924 & 200.5924				    \\
Tl & 81 & 203 & 29.524 &  & 0.054 &  & 3.33E-10 & 202.9723442 &  & 				    \\
Tl & 81 & 205 & 70.476 &  & 0.128 &  & 8.02E-10 & 204.9744275 &  & 				    \\
Tl & 81 &  & 100 & 0.181 &  & 1.13E-09 &  &  & 204.38333 & 204.38333			    \\
Pb$^{\wedge}$ & 82 & 204 & 1.9968 &  & 0.066 &  & 4.10E-10 & 203.9730436     \\
Pb & 82 & 206 & 18.5823 &  & 0.612 &  & 3.85E-09 & 205.9744653     \\
Pb & 82 & 207 & 20.5631 &  & 0.677 &  & 4.29E-09 & 206.9758969     \\
Pb & 82 & 208 & 58.8578 &  & 1.938 &  & 1.23E-08 & 207.976652 &  & 				    \\
Pb & 82 &  & 100 & 3.293 &  & 2.09E-08 &  &  & 207.3163 & 207.31887				    \\
Bi$^{\wedge}$ & 83 & 209 & 100 & 0.142 & 0.142 & 9.06E-10 & 9.06E-10 & 208.9803987 & 208.9804 & 208.98040     \\
Th$*$ & 90 & 232 & 100 & 0.0431 & 0.0431 & 3.06E-10 & 3.06E-10 & 232.0380553 & 232.03806 & 232.03806  \\
U$*$ & 92 & 234 & 0.0042 &  & 9.90E-07 &  & 7.09E-15 & 234.0409521     \\
U$*$ & 92 & 235 & 24.3016 &  & 0.0058 &  & 4.16E-11 & 235.0439299     \\
U$*$ & 92 & 238 & 75.6942 &  & 0.018 &  & 1.31E-10 & 238.0507882     \\
U & 92 &  & 100 & 0.0238 &  & 1.73E-10 &  &  & 238.02891 & 237.31991                            \\
\end{longtable}
\begin{tablenotes}
\small
\item $^{*}$  Table modified from Lodders (2020, 2021) where more references and details can be found. Atomic Masses are from Wang et al. (2021). Elements marked with $*$ involve long-lived radioactive nuclides with half-lives up to $10^{12}$ years and abundances are for $4.567$ Ga ago. Isotopes with half-lives above $10^{12}$ years (marked with a $\wedge$) can be considered as stable compared to the age of the solar system and are of interest for studies of double-beta decay. Isotopic compositions mainly adopted from Meija et al. (2016), except for H, C, N, O, and the noble gases. See Lodders (2020) for details and references.
\end{tablenotes}
\end{ThreePartTable}
